\documentclass[aps,notitlepage,floatfix,nofootinbib]{revtex4-1}

\usepackage{graphicx}
\usepackage{epstopdf}
\usepackage{caption}
\usepackage{bm}
\usepackage[latin1]{inputenc}
\usepackage{amsmath}
\usepackage{amsfonts}
\usepackage{subfigure}
\usepackage{amssymb}
\usepackage{makeidx}
\usepackage{color}
\usepackage{mathrsfs}
\RequirePackage[colorlinks,citecolor=blue,urlcolor=blue,linkcolor=magenta]{hyperref}

\graphicspath{{figures/}}

\begin{document}
\title{Cosmological dynamics of scalar fields with kinetic corrections: \\ Beyond the exponential potential}

\smallskip

\author{\bf~ Jibitesh~Dutta$^{1,2}$\footnote{jdutta29@gmail.com, jdutta@iucaa.ernet.in}, Wompherdeiki Khyllep$^3$\footnote{sjwomkhyllep@gmail.com}, Nicola Tamanini$^4$\footnote{nicola.tamanini@cea.fr}}

\smallskip

\affiliation{$^{1}$Mathematics Division, Department of Basic
Sciences and Social Sciences,~ North Eastern Hill University,~NEHU
Campus, Shillong - 793022 , Meghalaya, India}

\affiliation{$^{2}$ Visiting Associate, Inter University Centre
for Astronomy and Astrophysics, Pune, India}

\affiliation{$^3$Department of Mathematics, St. Anthony's College,
Shillong - 793001, Meghalaya, India}

\affiliation{$^4$Institut de Physique Th\'eorique, CEA-Saclay,
CNRS UMR 3681, Universit\'e Paris-Saclay, F-91191 Gif-sur-Yvette, France}

\date{\today}

\begin{abstract}
   We expand the dynamical systems investigation of cosmological scalar fields characterised by kinetic corrections presented in [N.~Tamanini, Phys.\ Rev.\ D {\bf 89} (2014) 083521].
   In particular we do not restrict the analysis to exponential potentials only, but we consider arbitrary scalar field potentials and derive general results regarding the corresponding cosmological dynamics.
   Two specific potentials are then used as examples to show how these models can be employed not only to describe dark energy, but also to achieve dynamical crossing of the phantom barrier at late times.
   Stability and viability issues at the classical level are also discussed.
\end{abstract}

\maketitle


\section{Introduction}

Dark energy is still one of the biggest mysteries of our universe.
Although its existence is by now observationally well confirmed \cite{Riess:1998cb,Perlmutter:1998np,Betoule:2014frx,Ade,Ade:2015xua,Amendola}, the nature and dynamics of this entity responsible for the present cosmic acceleration, is still unclear.
The easiest theoretical explanation comes from the modification of the Einstein field equations by the simple addition of a cosmological constant $\Lambda$.
The resulting $\Lambda$CDM model, accounting for both (cold) dark matter and dark energy, is well in agreement with the current observational data \cite{Betoule:2014frx,Ade,Ade:2015xua}, but it also suffers from unsolved theoretical prejudices, such as the so-called cosmological constant problem \cite{Weinberg:1988cp,Martin:2012bt} and the cosmic coincidence problem \cite{Zlatev}.

In order to solve these problems, or at least to avoid them, a plethora of alternative cosmological models have been proposed.
Almost all of them require the introduction of further dynamical degrees of freedom, either in the form of some new fields, or by modifying the underlying gravitational theory, namely general relativity (GR).
In the first case some new fields (or equivalently particles) beyond the Standard Model are postulated to appear in gravity's source sector (the right hand side of the Einstein equations), while in the second one the modified dynamics arises from new, often higher order, terms in the gravitational field itself (i.e.~in the left hand side of Einstein equations).

In both scenarios, the simplest models beyond $\Lambda$CDM assume the presence of a scalar field, which is either directly introduced in the matter sector or it stands out after some transformations and redefinitions have been performed, as for example in the case of $f(R)$ gravity \cite{Sotiriou:2008rp,DeFelice:2010aj}, which can be recast into Scalar-Tensor theories.
Scalar fields might in fact provide a sufficiently complex cosmological evolution, and being quite simple to handle from both a theoretical and conceptual point of view.
They are moreover well motivated by the low-energy limit of some well known high-energy theories; for example string theory.
For these reasons scalar fields are commonly employed not only to describe dark energy (see e.g.~\cite{Copeland:2006wr,Tsujikawa:2013fta} for reviews), but also to characterize inflation \cite{Liddle:2000cg}, dark matter \cite{Magana:2012ph} and even unified dark sector models \cite{Bertacca:2010ct}.

The most general non-minimally coupled scalar field that gives second order field equations, is defined by the so-called Horndeski Lagrangian \cite{Horndeski:1974wa} where higher order derivatives of the scalar field $\phi$ appear.
The first and simplest term of this Lagrangian, which arbitrarily depends on both $\phi$ and its kinetic energy, has often been used to build alternative scalar field models of dark energy, which have collectively been named $k$-essence theories \cite{ArmendarizPicon:2000ah}.
Within this class of theories it is possible to obtain a more complicated cosmological dynamics which cannot be easily produced with a canonical scalar field.
For example an interesting possibility is the dynamical crossing of the phantom barrier, though one must be aware of the instabilities that arise in such situations \cite{Vikman:2004dc,Zhao:2005vj,Caldwell:2005ai}.

In this paper we extend the analysis performed in \cite{Nicola}, where two particular $k$-essence models have been studied using dynamical systems techniques.
In \cite{Nicola} the scalar field was characterized by an exponential potential, whose evolution is usually easier to investigate with dynamical systems methods \cite{Copeland:1997et,TamaniniPhDthesis}.
In this work instead we analyse the same non-canonical cosmological models considered in \cite{Nicola} by assuming general self-interacting scalar field potentials $V(\phi)$.
Moreover, in order to get a deeper insight into their cosmological dynamics (especially regarding non-hyperbolic critical points), we also propose two concrete potentials as examples: the hyperbolic potential $V = V_0~\rm sinh^{- \alpha}(\lambda\phi)$ and the inverse power-law potential $V=\frac{M^{4+n}}{\phi^n}$.

Dynamical systems are extremely useful for analysing the complete
evolution of any background cosmological model (see
\cite{Garcia-Salcedo:2015ora,Boehmer:2014vea} for introductions to
the subject). Their applications to cosmology have in fact a long
history and a somehow flourishing literature, especially after the
discovery of dark energy in 1998. For some earlier and more
mathematical results we refer the reader to \cite{Ellis,Coley},
while an extended review of applications to dark energy models,
including modified gravity theories, can be found in
\cite{TamaniniPhDthesis}. Furthermore a sample of some recent
dynamical systems work in cosmology might be given by the
following references: quintessence
\cite{Gosenca:2015qha,Alho:2015ila,Paliathanasis:2015gga,Mahata:2015sry,Qi:2015pqa},
interacting dark energy
\cite{Szydlowski:2015rga,Singh:2015rqa,Shahalam:2015sja,Landim:2015uda,Fay:2015sjt},
chameleon theories \cite{Roy:2015cna}, Brans-Dicke theory
\cite{Garcia-Salcedo:2015naa,Cid:2015pja}, $f(R)$ gravity
\cite{Kandhai:2015pyr,Carloni:2015jla}, hybrid metric-Palatini
gravity \cite{Carloni:2015bua,Tamanini:2013ltp}, non-minimally
coupled scalar fields
\cite{Sakstein:2015jca,Hrycyna:2015eta,Bhattacharya:2015wlz},
Higgs dark energy \cite{Rinaldi:2014yta,Rinaldi:2015iza},
tachyonic dark energy \cite{Landim:2015poa,Mahata:2015lja},
braneworld scenarios \cite{Biswas:2015zka,Dutta:2015jaq}, modified
teleparallel theories of gravity
\cite{Skugoreva:2014ena,Carloni:2015lsa,Biswas:2015cva,Bahamonde:2015hza}
and Scalar-Fluid theories \cite{Boehmer:2015kta,Boehmer:2015sha}.

The rest of the paper is organized as follows.
In Sec.~\ref{basics} we discuss the generalized class of scalar field theories that will be studied in this paper, we present the resulting cosmological equations and we introduce the dimensionless normalized variables that will be used for the dynamical systems analysis.
In Sec.~\ref{sec:sqare_kc} we consider square kinetic corrections to the canonical scalar field Lagrangian and we investigate the corresponding cosmological equations employing dynamical systems techniques.
In Sec.~\ref{sec:sqr_r_kc} we repeat the same analysis for square root kinetic corrections.
In both sections the two potentials mentioned above are used as examples to better clarify the possible cosmological dynamics arising from these models, especially regarding non-hyperbolic critical points.
Finally in Sec.~\ref{sec:conclusions} we draw our conclusions.

{\it Notation}: We consider units where $8\pi G=c=1$ and use the $(-,+,+,+)$ signature convention for the metric tensor.

\section{Non canonical scalar field models and basic cosmological equations}
\label{basics}

The total action of a scalar field minimally coupled to gravity is given by
\begin{equation}\label{eq1}
S=\int d^4x\sqrt{-g}\left[\frac{R}{2}+\mathcal{L}_{\phi}+\mathcal{L}_m\right] \,,
\end{equation}
where $g$ is the determinant of the metric $g_{\mu\nu}$, $R$ is the Ricci scalar, $\mathcal{L}_m$ denotes the matter Lagrangian and $\mathcal{L}_{\phi}$ denotes the scalar field Lagrangian.
In this paper we focus on the following scalar field Lagrangian
\begin{equation}\label{eq3}
\mathcal{L}_{\phi}=V f(B) \,,
\end{equation}
where $V(\phi)$ denotes an arbitrary potential for the scalar
field and $f$ is an arbitrary function of
\begin{equation}\label{eq4}
B=\frac{X}{V} \,, \quad\text{with}\quad X=-\frac{1}{2}g^{\mu\nu}\partial_{\mu}\phi\partial_{\nu}\phi \,.
\end{equation}
This type of scalar field has already been considered in \cite{Tsujikawa:2004dp,Piazza:2004df,Nicola} as an alternative model of dark energy leading to late time attractor solutions and it is well motivated from high-energy physics considerations; see e.g.~\cite{Copeland:2006wr} Sec.~V.B or \cite{DEbook} Chapter~8 and references therein.
In more details the low-energy effective string theory generates higher-order derivative terms in the dilaton (scalar) field coming from the string length scale and loop corrections \cite{Gasperini:2002bn}.
Such higher-order corrections in the scalar field Lagrangian have then first been used for building alternative models of inflation \cite{ArmendarizPicon:1999rj}, and then to characterise dark energy \cite{Copeland:2006wr,DEbook}, in particular as a dilatonic ghost condensate \cite{Piazza:2004df}.

Astronomical observations favour a spatially flat, homogeneous and isotropic Friedmann-Robertson-Walker (FRW) universe \cite{Miller,Ade,Ade:2015xua}, well described by the metric
\begin{equation}\label{eq2}
ds^2=-dt^2+a^2(t)\left(dx^2+dy^2+dz^2\right) ,
\end{equation}
where $a(t)$ is the scale factor, $t$ is the physical time and $x,y,z$ are Cartesian coordinates.
Varying action (\ref{eq1}) with respect to the metric tensor $g_{\mu\nu}$ and using the FRW metric \eqref{eq2}, we obtain the Friedmann equation
\begin{equation}\label{eq5}
3H^2=\rho_m-V f+\frac{\partial f}{\partial B}\dot{\phi}^2 \,,
\end{equation}
where $H=\dot{a}/a$ is the Hubble parameter and $\rho_m$ is the energy density of the matter perfect fluid.
On the other hand, the variation of action (\ref{eq1}) with respect to $\phi$ yields the following modified cosmological Klein-Gordon equations
\begin{equation}\label{eq6}
\left(\frac{\partial f}{\partial B}+2 B \frac{\partial^2 f}{\partial B^2}\right)\ddot{\phi}+3H\dot{\phi}\frac{\partial f}{\partial B}-\left(f-\frac{\partial f}{\partial B}+2 B^2 \frac{\partial^2 f}{\partial B^2}\right)\frac{dV}{d{\phi}}=0 \,.
\end{equation}

The energy density and pressure of the scalar field are given by
\begin{eqnarray}
\rho_{\phi}&=&2 X \frac{\partial f}{\partial B}-V f \,,\label{eq9}\\
p_{\phi}&=&\mathcal{L}_{\phi}=V f \,.\label{eq10}
\end{eqnarray}
Using this notation one can write the conservation equation of the scalar field and the one of the matter fluid as
\begin{eqnarray}
\dot{\rho}_{\phi}+3H(\rho_{\phi}+p_{\phi})&=&0 \,, \label{eq7}\\
\dot{\rho}_m+3H(\rho_m+p_m)&=&0\label{eq8} \,,
\end{eqnarray}
where $\rho_m$ and $p_m$ denote the energy density and pressure of the perfect fluid with a barotropic equation of state (EoS) $w$ given by $p_m=w\rho_m$ $(-1\leq w \leq 1)$.
The EoS parameter of the scalar field is given by
\begin{equation}\label{eq11}
w_{\phi}=\frac{p_{\phi}}{\rho_{\phi}}=\left(2\frac{X}{V}\frac{1}{f}\frac{\partial f}{\partial B}-1\right)^{-1} \,.
\end{equation}
Finally another important quantity needed to determine the stability of any cosmological model at the classical level is the adiabatic speed of sound $C_s^2$ defined by
\begin{equation}\label{eq12}
C_s^2=\frac{\partial p_{\phi}/\partial X}{\partial \rho_{\phi}/\partial X}=\left(1+2\frac{X}{V}\frac{\partial^2f/\partial X^2}{\partial f/\partial X}\right)^{-1} \,.
\end{equation}

Following \cite{Nicola}, we introduce dimensionless phase space variables, in order to write the above cosmological equations as an autonomous system of equations:
\begin{equation}\label{eq13}
x=\frac{\dot{\phi}}{\sqrt{6} H}\,,~~~~ y=\frac{\sqrt{V}}{\sqrt{3}H}\,,~~~~ z=\frac{\sqrt{\rho_m}}{\sqrt{3}H}\,,~~~~s=-\frac{1}{V}\frac{dV}{d\phi} \,.
\end{equation}
The variable $s$, first employed in \cite{Steinhardt:1999nw,delaMacorra:1999ff,Ng:2001hs}, is commonly introduced in order to study the dynamics for arbitrary self interacting scalar field potentials \cite{TamaniniPhDthesis,Zhou:2007xp,Fang:2008fw,Matos:2009hf,UrenaLopez:2011ur}. In terms of the dimensionless variables (\ref{eq13}), the Friedmann equation (\ref{eq5}) becomes
\begin{equation}\label{eq14}
1=z^2-y^2f+2 x^2\frac{\partial f}{\partial B} \,,
\end{equation}
while the effective EoS parameter is defined as
\begin{equation}\label{eq15}
w_{\rm eff}=\frac{p_{\phi}+p_m}{\rho_{\phi}+\rho_m}=w+(w+1)y^2 f-2 w x^2 \frac{\partial f}{\partial B} \,.
\end{equation}
The accelerated expansion of the universe is attained whenever the condition $w_{\rm eff}<-\frac{1}{3}$ is realised.

In order to study the cosmological dynamics of Eqs.~\eqref{eq5}--\eqref{eq8} in more detail, one needs first to specify the form of the function $f$.
One can easily verify that the choice $f(B)=B-1$ corresponds to a canonical scalar field with Lagrangian $\mathcal{L}_{\phi}=X+V$.
In this paper we consider general scalar field potentials $V(\phi)$ for extended Lagrangians characterized by the function
\begin{equation}\label{eq:001}
    f(B) = B-1+\xi B^n \,,
\end{equation}
Such choices generally describe higher or lower order (depending on $n$ being bigger or smaller than one) kinetic corrections to the standard canonical scalar field Lagrangian, reducing to the latter one in the limit $\xi \rightarrow 0$.
In what follows we focus on two different choices of $f$: $n=2$ and $n=1/2$, corresponding to square and square root kinetic corrections to the canonical case.
The dynamics of these two models has been studied in \cite{Nicola} for scalar field potentials of the exponential kind, but different types of potentials have never been investigated before.
On the other hand, regarding the canonical scalar field, the corresponding analysis beyond the exponential potential has already been done in \cite{Zhou:2007xp,Fang:2008fw,Matos:2009hf,UrenaLopez:2011ur} (see also \cite{TamaniniPhDthesis}, Sec.~4.4), and will thus not be considered here.


\section{Square Kinetic Corrections}
\label{sec:sqare_kc}

We first analyse the cosmological dynamics arising from square kinetic corrections to the canonical scalar field Lagrangian, i.e.,~the case $n=2$.
The scalar field Lagrangian is specifically given by
\begin{equation}\label{eq16}
\mathcal{L}_{\phi}=X-V+\xi \frac{X^2}{V} \,,
\end{equation}
where the parameter $\xi$ can be any real number.
Note that when $V\gg X$ the Lagrangian reduces to the one of a canonical scalar field dominated by its potential energy.
At cosmological scales this usually happens at late times, implying $\rho_\phi \approx V(\phi)$ and thus that accelerated expansion can effectively be driven by the scalar field potential.

Using the Lagrangian \eqref{eq16}, the Friedmann equation (\ref{eq14}) becomes
\begin{equation}\label{eq17}
1=x^2+y^2+3\xi\frac{x^4}{y^2}+z^2 \,.
\end{equation}
This is a constraint that the dimensionless variables \eqref{eq13} must always satisfy.
Using \eqref{eq17} we can thus substitute $z^2$ in terms of the other variables whenever it appears in the equations that follow, effectively reducing the dimension of the system, i.e.~the number of variables, from four to three.
Moreover, given the physical requirement $\rho_m \geq 0$, the constraint $0 \leq z^2\leq 1$ must hold.
From \eqref{eq17} one can thus bound the phase space, at least on $(x,y)$-planes, as
\begin{equation}
    x^2+y^2+3\xi\frac{x^4}{y^2} = 1-z^2 \leq 1 \,.
    \label{eq:constr_sq}
\end{equation}

In terms of the dimensionless variables (\ref{eq13}) the following relevant cosmological parameters viz., the relative scalar field energy density parameter, the relative energy density parameter of matter, the EoS parameter of scalar field, the effective EoS parameter and deceleration parameter, are respectively given by
\begin{eqnarray}
\Omega_{\phi}&=&\frac{\rho_{\phi}}{3H^2}=x^2+y^2+3\xi\frac{x^4}{y^2} \,, \label{eq18}\\
\Omega_{m}&=&\frac{\rho}{3H^2}=1-x^2-y^2-3\xi\frac{x^4}{y^2}\label{eq19} \,, \\
w_{\phi}&=&\frac{(x^2-y^2)y^2+\xi x^4}{(x^2+y^2)y^2+3\xi x^4}\label{eq20} \,, \\
w_{\rm eff}&=&w-(w-1)x^2-(w+1)y^2-\xi (3w-1)\frac{x^4}{y^2}\label{eq21} \,, \\
q&=&-1-\frac{\dot{H}}{H^2}=-1-\frac{3}{2y^2}\left[\xi(3w-1)x^4+(w-1)x^2y^2+(w+1)y^2(y^2-1)\right]\label{eq21a} \,.
\end{eqnarray}
When $y$ dominates over $x$ and $\sigma$, i.e.~when $V\gg X, \rho$ and one has $y \approx 1$ and $x, \sigma \approx 0$, one finds $w_{\rm eff} \approx w_\phi \approx -1$ and the universe is well described by an accelerating de Sitter solution.
On the other hand when $X\gg V$ we have $w_\phi \approx 1/3$, implying that the scalar field effectively describes a relativistic fluid.
Finally the adiabatic speed of sound is given by
\begin{equation}\label{eq22}
C_s^2=\frac{y^2+2\xi x^2}{y^2+6\xi x^2} \,.
\end{equation}
We see that $C_s^2>0$ and $\Omega_{\phi}>0$ whenever $\xi>0$.
In order for this model to be physically viable we thus only consider positive values of $\xi$.
Moreover, in the numerical examples that follow, we estimate the final state of the universe in such a way that it matches the present observational data ($\Omega_m=0.27,~q=-0.53$) \cite{Goistri}.
This corresponds to points in the phase space where
\begin{equation}\label{22b}
x_0=\pm 0.142,~~y_0=\pm 0.841 \,.
\end{equation}

Using the dimensionless variables (\ref{eq13}), the cosmological equations can be recast into the following autonomous system of equations
\begin{eqnarray}
x'&=&\frac{1}{2y^2\,(6x^2\xi+y^2)}\Big[18\xi^2(1-3w)\,x^7-3x\,y^4\,\left(6\,\xi\,x^2\,(w+1)+(w-1)\,(x^2-1)\right)\nonumber \\&&~{}\left.+3\,\xi\,x^3\,y^2\,\left((7-9\,w)\,x^2+6\,w-\sqrt{6}\,s\,x+2\right)+y^6\,\left(\sqrt{6}\,s-3(w+1)\,x\right)\right] \,, \label{eq23}\\
y'&=&\frac{1}{2y}\left[3y^2(w+1)(1-y^2)+3\xi(1-3w)x^4-xy^2\left(\sqrt{6}\,s+3(w-1)x\right)\right] \,, \label{eq24}\\
s'&=&-\sqrt{6}\,x\,s^2 g(s) \,, \label{eq25}
\end{eqnarray}
where we have defined $g(s)=\Gamma(s)-1$ and
\begin{equation}
    \Gamma= V\ \frac{d^2 V}{d\phi^2} \left( \frac{d V}{d \phi} \right)^{-2}  \,. 
\end{equation}
In Eqs.~\eqref{eq23}--\eqref{eq25}, and in the following equations, a prime denotes differentiation with respect to the number of $e$-folds $N$, defined such that $dN = H dt$.
We  note that the system (\ref{eq23})-(\ref{eq25}) is invariant under the transformation $y\rightarrow-y$. So we will focus only on positive values of $y$, since the dynamics on the positive $y$ region is a mirror image of the negative $y$ region.
Moreover, as mentioned above, taking into account the physical condition $\rho_m\geq 0$, one must assume $z^2\geq 0$ which leads to the constraint
\begin{equation}\label{eq26}
0\leq x^2+y^2+3\xi\frac{x^4}{y^2}\leq 1 \,.
\end{equation}
Hence, the 3D phase space of the system (\ref{eq23})-(\ref{eq25}) is given by
\begin{equation}\label{eq27}
\Psi=\left\lbrace (x,y)\in \mathbb{R}^2: 0\leq x^2+y^2+3\xi\frac{x^4}{y^2}\leq 1\right\rbrace \times \left\lbrace s \in \mathbb{R}\right\rbrace \,.
\end{equation}
\begin{center}
\begin{table}[t]
\caption{Critical points and corresponding cosmological parameters of the system (\ref{eq23})-(\ref{eq25}) for a generic
scalar field potential.}
\begin{center}
\begin{tabular}{c c c c  c c c c}
\hline\hline
Point&$~~~x~~~$&$~~~y~~~$&$~~~s~~~$& ~~~Existence~~~&$~~~\Omega_\phi~~~$&$~~~\omega_{\rm eff}~~~$&Acceleration \\ \hline\\
$A_1$&  $0$   &$0$    &$s$      &Always        & $0$             & $w$            &No\\ [1.5ex]
$A_2$&  $0$    &$1$   &$0$    &Always   & $1$             & $-1$            &Always\\ [1.5ex]
$A_3$&  $\frac{\sqrt{6}(w+1)}{2s_*}$    &$\frac{\sqrt{3 \xi (3w-1)}(w+1)}{s_*\sqrt{(-w+1+\delta)}}$   &$s_*$    & Fig.\ref{A3A4s} & $\frac{3(w+1)^2(4\xi (1-3w)+1-w+\delta)}{s_*^2(3w-1)(1-w+\delta)}$             & $w$            &No\\ [1.5ex]
$A_4$&  $x_4$    &$y_4$   &$s_*$    &Fig.\ref{A3A4s}  
 &    $1$          &     App.        &Fig.\ref{A3A4s}\\ [1.5ex]
\hline\hline
\end{tabular}\label{Tab1}
\begin{center}
$\delta=\sqrt{4 w^2-(3w-1)(w+1)(4\xi+1)}$\\[1.5ex]

$x_4=\frac{\Pi^\frac{2}{3}+4s_*\Pi^\frac{1}{3}-36\xi(s_*^2+4)+7s_*^2-36}{3\sqrt{6}(4\xi+1)\Pi^\frac{1}{3}}$,~~ $y_4=\frac{1}{\sqrt{6}}\left[3(x_4^2+1)-\sqrt{6}s_*x_4+\sqrt{36 \xi x_4^4+\left(3x_4^2-\sqrt{6}s_* x_4+3\right)^2}\right]^{\frac{1}{2}}$\\[1.5ex]

$\Pi=54(48\xi^2+8\xi-1)s_*+(10-216\xi)s_*^3+9(4\xi+1)\left[36\xi(s_*^6-12s_*^4+24s_*^2+64)+5148\xi^2s_*^2-(s_*^2-6)^2(3s_*^2-16)\right]^\frac{1}{2}$
\end{center}
\end{center}
\end{table}
\end{center}
\begin{center}
\begin{table}[t]
\caption{Eigenvalues of critical points listed in Table~\ref{Tab1}.}
\begin{center}
\begin{tabular}{c c c c c}
\hline\hline
Point  &  $E_1$   &$E_2$     &$E_3$  &Stability  \\ \hline\\
$A_1$&  $\frac{3}{2}(w-1)$   &$\frac{3}{2}(w+1)$    &$0$& saddle
\\ [1.5ex]
$A_2$&  $-3(1+w)$    &$-3$   &$0$& stable/saddle    \\ [1.5ex]
$A_3$&  $e_+$    &$e_-$   &$-\frac{3\sqrt{6}}{2 s_*}\,dg(s_*)$& stable/saddle   \\ [1.5ex]
$A_4$&  App.   &App.  &App.& Fig. \ref{A3A4s}   \\ [1.5ex]
\hline\hline
\end{tabular}\label{Tab2}
\end{center}
\begin{center}
\begin{tiny}
Note: For point $A_3$, to avoid lengthy expressions, eigenvalues are calculated by considering the case $w=0$. For point $A_4$, eigenvalues are solved numerically and are not given in Table~\ref{Tab2} due to their lengthy expressions, which are provided in the Appendix.
\end{tiny}
\end{center}
\begin{center}
$e_{\pm}=-\frac{3}{4}\left(1\pm{\frac {\sqrt{16\,\sqrt {4\,\xi+1}{s_*}^{2}-60\,{s_*}^{2}\xi-19\,{s_*}^{2}+72\,\xi \left(
1-4\,\sqrt {4\,\xi+1} \right) -120\,\xi-12
}}{ \left( 36\,\xi
+5 \right) s_*}}\right)$
\end{center}
\end{table}
\end{center}
\begin{figure}
\centering
\subfigure[]{%
\includegraphics[width=8cm,height=6cm]{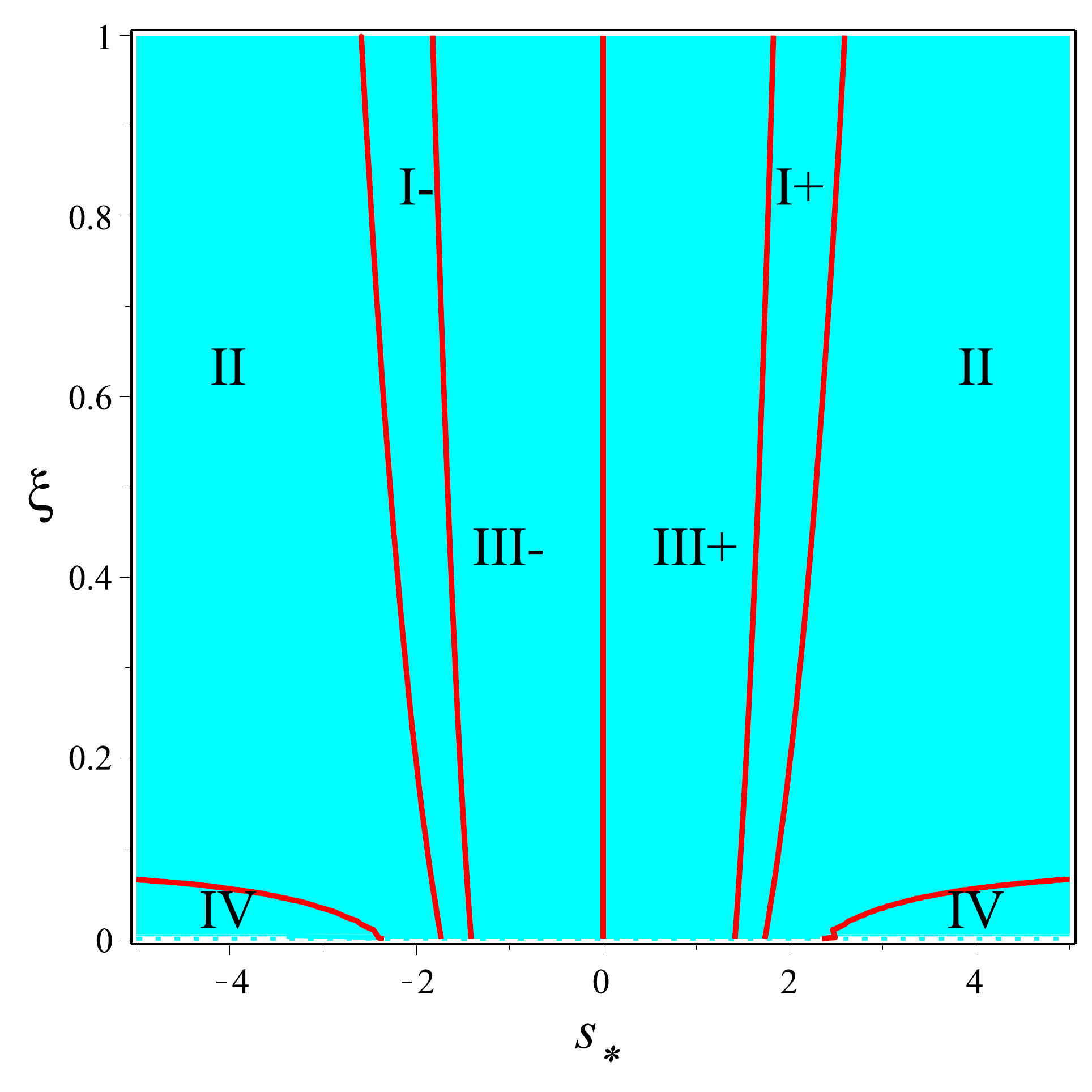}\label{existA3A4s}}
\qquad
\subfigure[]{%
\includegraphics[width=8cm,height=6cm]{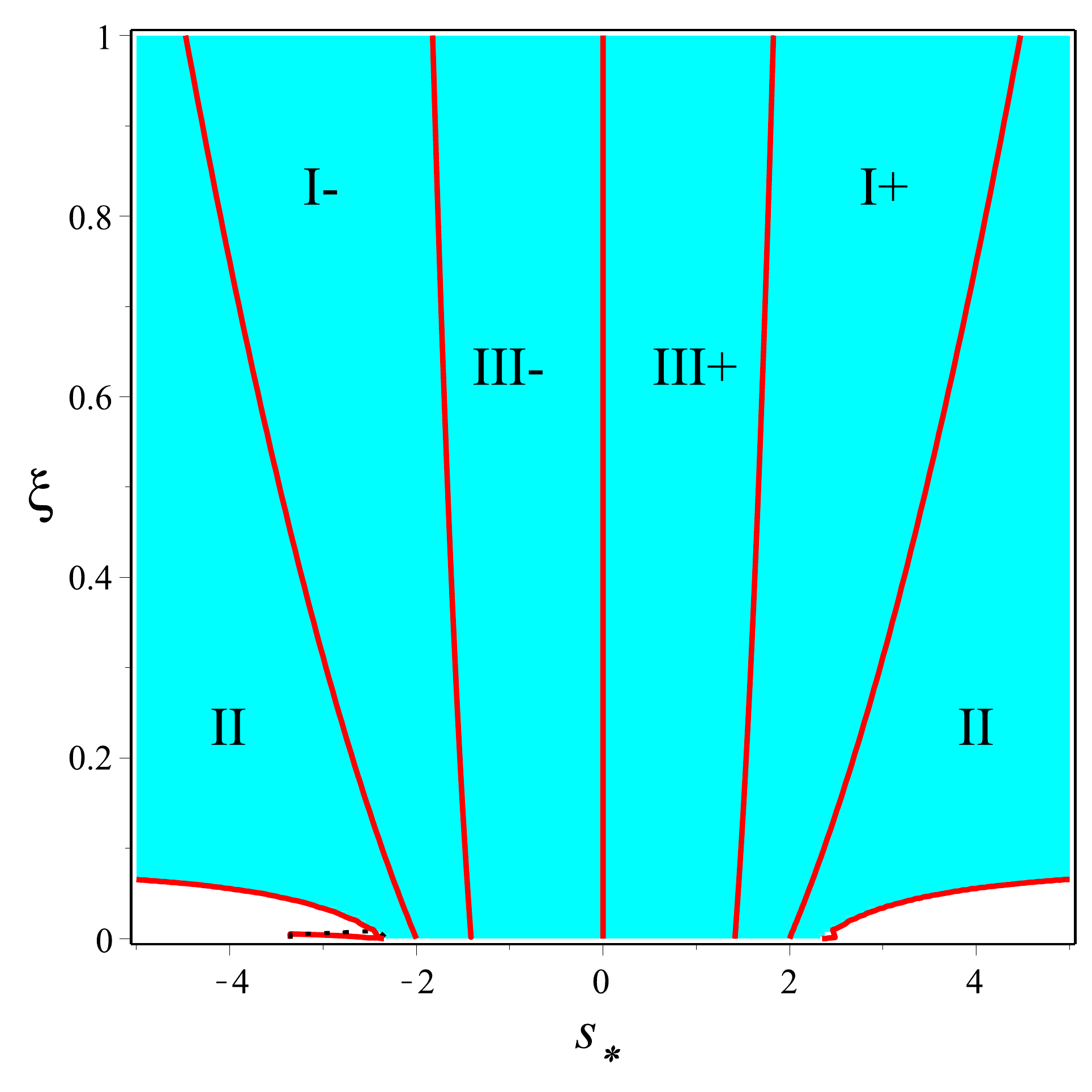}\label{existA3A41by3s}}
\caption{Existence of points~$A_3$ and $A_4$ in the $(s_*,\xi)$ parameter space. Panel (a) corresponds to the case $w=0$ and panel (b)
corresponds to the case $w=\frac{1}{3}$. In the $w=0$ case, regions II and IV denotes the existence region of point $A_3$, while in the $w=\frac{1}{3}$ case point~$A_3$ does not exists at all. In both panels, regions I+, I$-$, II, III+ and III$-$ corresponds to the region of existence of point $A_4$; regions III+ and III- denote values of $s_*$ and $\xi$ for which point~$A_4$ describes an accelerated solution; whereas its stability is attained only in regions I+ and III+ for potentials where $dg(s_*)>0$ and in regions I$-$ and III$-$ for potentials where $dg(s_*)<0$. Here distinct numbered regions are separated by red solid curves.}
\label{A3A4s}
\end{figure}

The critical points of the system (\ref{eq23})-(\ref{eq25}) along with their relevant cosmological parameters are summarised in Table \ref{Tab1}.
The eigenvalues of the corresponding Jacobian matrices are instead given in Table \ref{Tab2}.
In all cases, $s_*$ represents a solution of $g(s)=0$.
The critical point $A_1$ exists for any arbitrary potential ($s$ can take any value), whereas point $A_2$ represents a solution where the scalar field potential is effectively constant ($s=0$).
The existence of points $A_3$ and $A_4$ depends on the specific potential under consideration (note also that there are as many points $A_3$ and $A_4$ as many solutions $s_*$ of $g(s)=0$).
The critical point $A_1$ does not formally belong to the phase space since $y=0$ fails to satisfy the constraint \eqref{eq26}.
However, by converting $x,~y$ to polar coordinates, one can actually verify that all trajectories in the phase space provide a well defined limit as they approach the origin.
For this reason in what follows we will consider point $A_1$ as belonging to the phase space.

Critical points $A_1$ and $A_2$  are non-hyperbolic critical
points, as one can realise looking at the corresponding vanishing
eigenvalues in Table~\ref{Tab2}. Point $A_1$ is a saddle
point since its two non vanishing eigenvalues (excluding the case
$w=-1$) always appear with opposite sign. The stability of point
$A_2$ is instead provided by the attracting nature along the
direction individuated by the eigenvector corresponding to the
vanishing eigenvalue. In order to determine the stability of this
point analytically one should apply the centre manifold theorem to
find the dynamics along its centre manifold (see
e.g.~\cite{TamaniniPhDthesis,Boehmer:2011tp}). However this can
also be found numerically, using some particular
perturbation techniques, once a specific potential has been
selected.
These numerical techniques turn out to be especially useful to study increasingly complicated non-hyperbolic critical points, whose stability cannot easily be addressed analytically, and have been extensively employed in some recent literature \cite{Roy:2015cna,Dutta:2015jaq, dutta2016}. Since this is
somehow simpler to do in practice, we postpone the stability
analysis of point $A_2$ to each particular example we consider
below.

Finally, due to the complicated expressions of critical points
$A_3$ and $A_4$, the complete analysis of their stability can only
be performed once a specific potential has been chosen.
Nevertheless we are still able
to draw some conclusions, by treating $s_*$ as a parameter, even without specifying a scalar field
potential $V$.
We will focus on the most physically interesting cases $w=0$ and $w=1/3$, for which the existence and stability regions in the $(\xi, s_*)$ parameter space have been drawn in Fig.~\ref{A3A4s}.

First point $A_3$ corresponds to an un-accelerated
scaling solution with $w_{\rm eff}=w$, and we find that it
does not exist when $w=\frac{1}{3}$, irrespectively of the value
of $s_*$, since it does not satisfy the constraint (\ref{eq26}). Moreover the
eigenvalues $E_1,~E_2$ are always negative within the region of
existence of point $A_3$. Therefore, point $A_3$ is a stable node
if $s_*dg(s_*)>0$ ($dg(s_*)$ is the derivative of $g(s)$ evaluated
at $s_*$) otherwise it is a saddle point, unless $dg(s_*)=0$ in
which case it is a non-hyperbolic critical point and its stability
can only be determined using the centre manifold theorem. Critical
point $A_4$ corresponds instead to a scalar field dominated solution. Its
eigenvalues are not listed in Table~\ref{Tab2} due to their
lengths, but are provided in the Appendix.
Moreover, due to their
complicated expressions, we can only check the stability of
point~$A_4$ numerically by choosing different values of $\xi$ and
$s_*$ as shown in Fig.~\ref{A3A4s}. From such
analysis we generally find that point $A_4$ can be a late time
accelerated attractor or a saddle point, depending on the values
of $\xi$, $s_*$ and $dg(s_*)$.
However since
this strongly depends, through $s_*$ and $dg(s_*)$, on the particular scalar
field potential under study, it would not be useful and neither
instructive to further investigate the stability of point~$A_4$
for a general potential. We thus independently repeat this analysis in the examples that follow, where a
specific scalar field potential will be assigned.

The general results obtained in this section are mathematically interesting since the properties of the phase space can be studied without specifying the form of the scalar field potential.
Nevertheless in order to find some physical applications of these models and better investigate their cosmological dynamics, one is forced to assume a particular potential.
In the remaining part of this section we thus choose two specific examples and analyse the resulting dynamics in detail.

\subsection{Example 1: $V=V_0 \sinh^{-\alpha}(\lambda\phi)$}

In this section we consider the potential
\begin{equation}\label{pot1}
    V(\phi) = V_0 \sinh^{-\alpha}(\lambda\phi) \,,
\end{equation}
where $V_0$ and $\lambda$ are two constants of suitable dimensions, while $\alpha$ is a dimensionless parameter.
This potential was first studied in \cite{Sahni}, where it has been shown that cosmological tracker solutions arise for quintessence.
For a canonical scalar field, it has then been studied employing dynamical systems techniques in \cite{Fang:2008fw,Roy:2013wqa}, while
the cosmological evolution of some alternative cosmological models using this potential has been investigated in \cite{Leyva}.

\begin{figure}
\centering
\subfigure[]{%
\includegraphics[width=5cm,height=3.75cm]{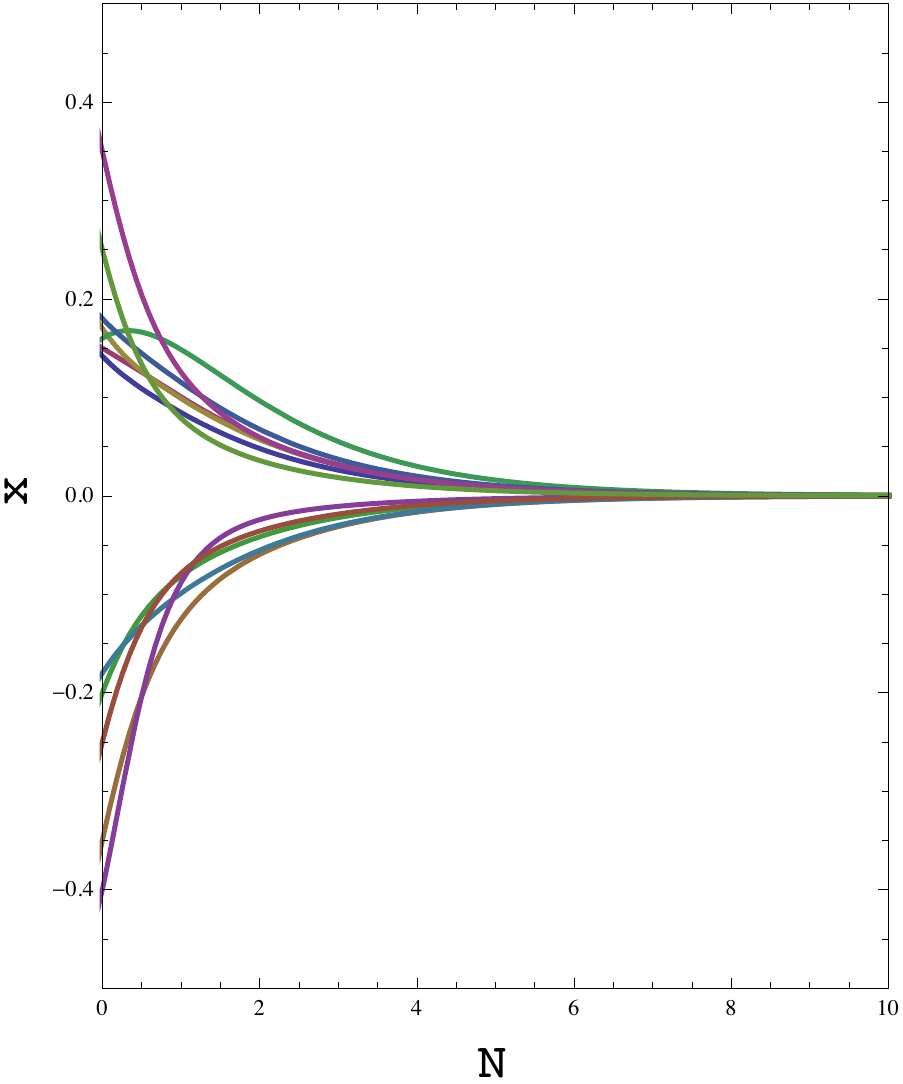}\label{fig1}}
\qquad
\subfigure[]{%
\includegraphics[width=5cm,height=3.75cm]{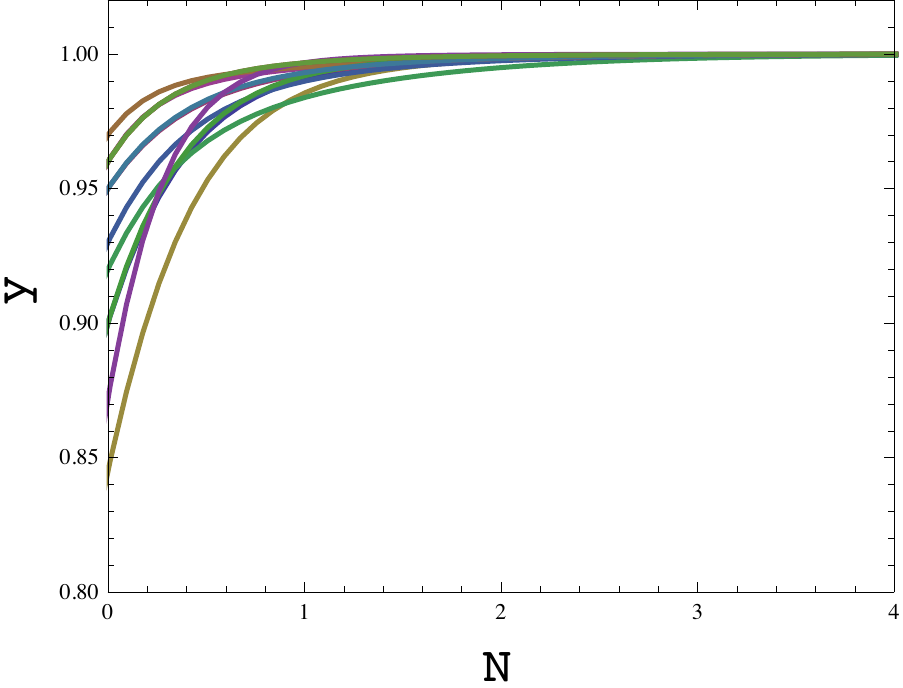}\label{fig2}}
\qquad
\subfigure[]{%
\includegraphics[width=5cm,height=3.75cm]{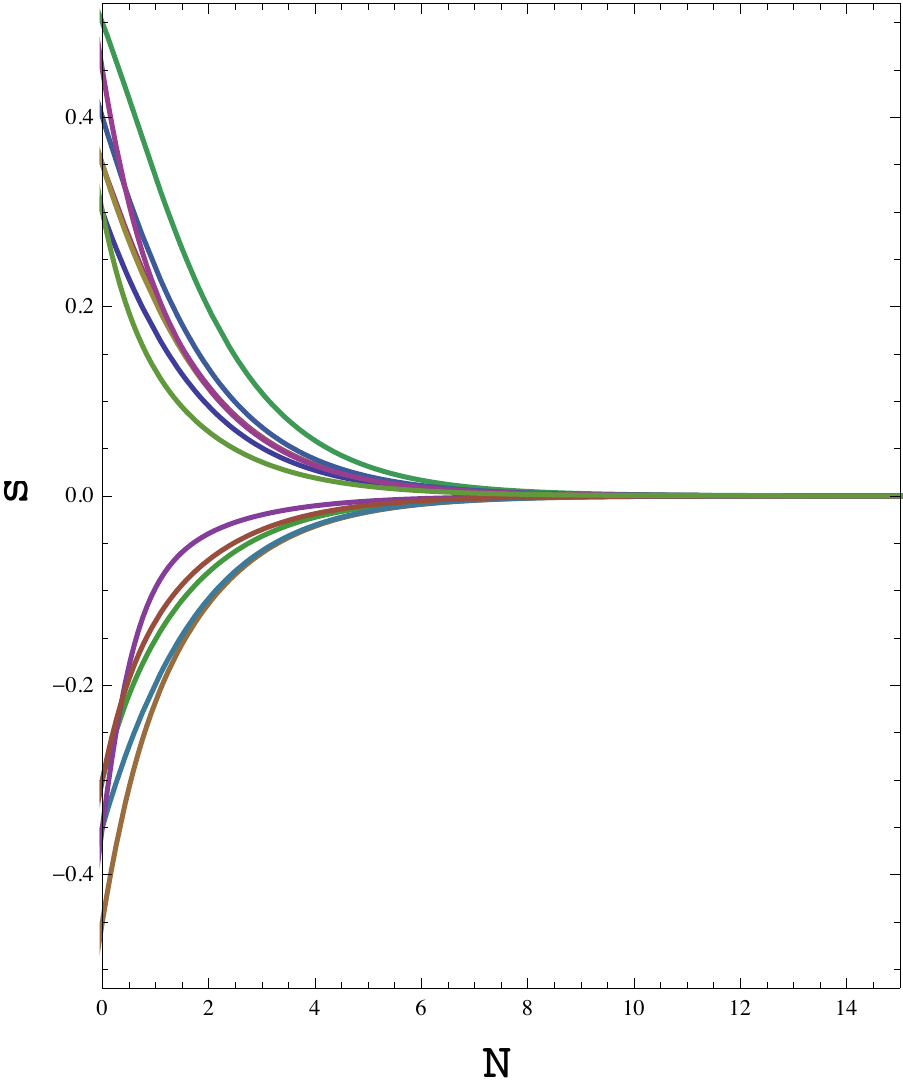}\label{fig3}}
\caption{(a).~Time evolution of $x$-projections of trajectories approaching point $A_2$; (b).~Time evolution of $y$-projections of trajectories approaching point $A_2$; (c).~Time evolution of $s$-projections of trajectories approaching point $A_2$. Here we have considered the potential \eqref{pot1} with $\xi=1$, $\alpha=-2$, $\lambda=0.5$ and $w=0$.}
\end{figure}
\begin{figure}
\centering
\subfigure[]{%
\includegraphics[width=8cm,height=6cm]{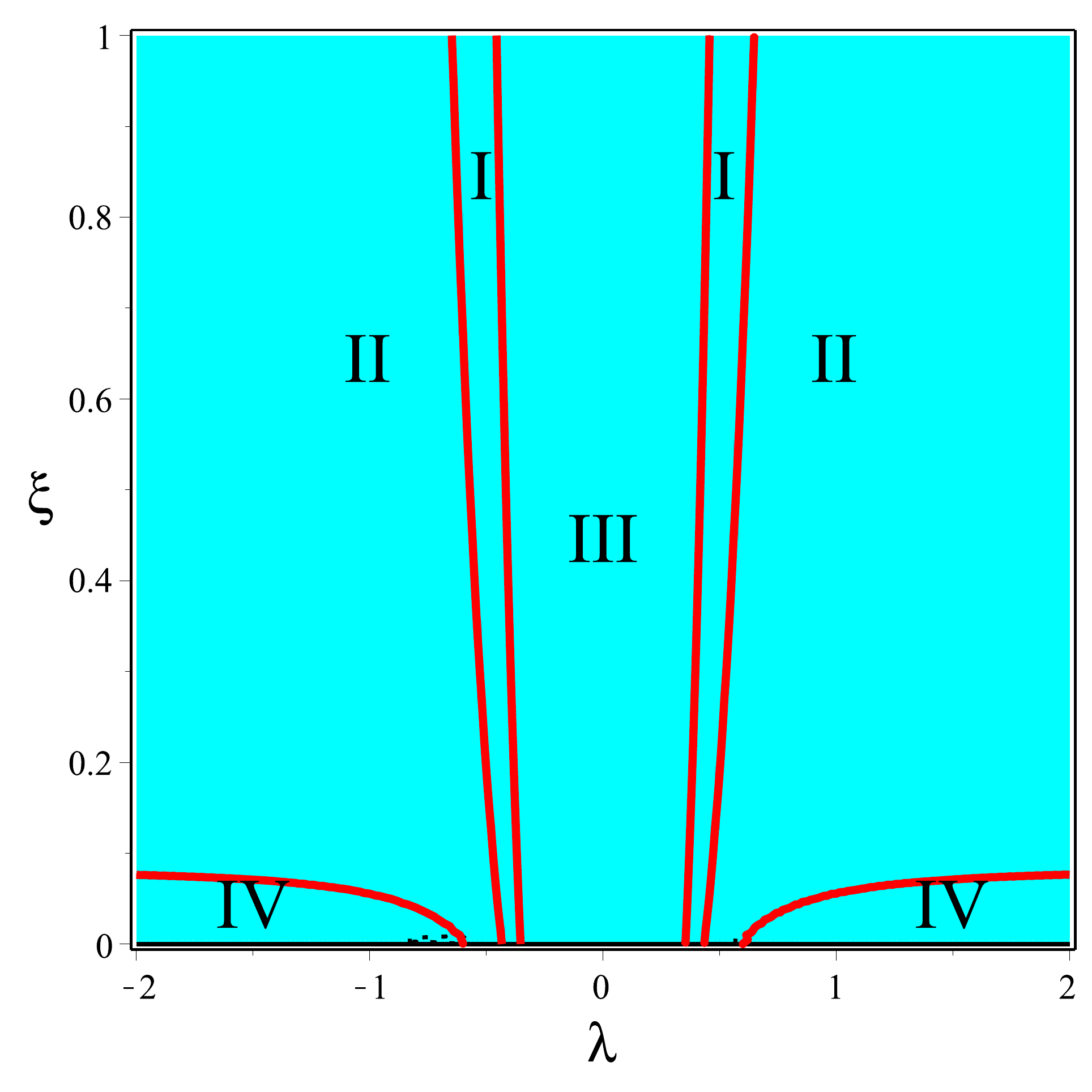}\label{existA3A4}}
\qquad
\subfigure[]{%
\includegraphics[width=8cm,height=6cm]{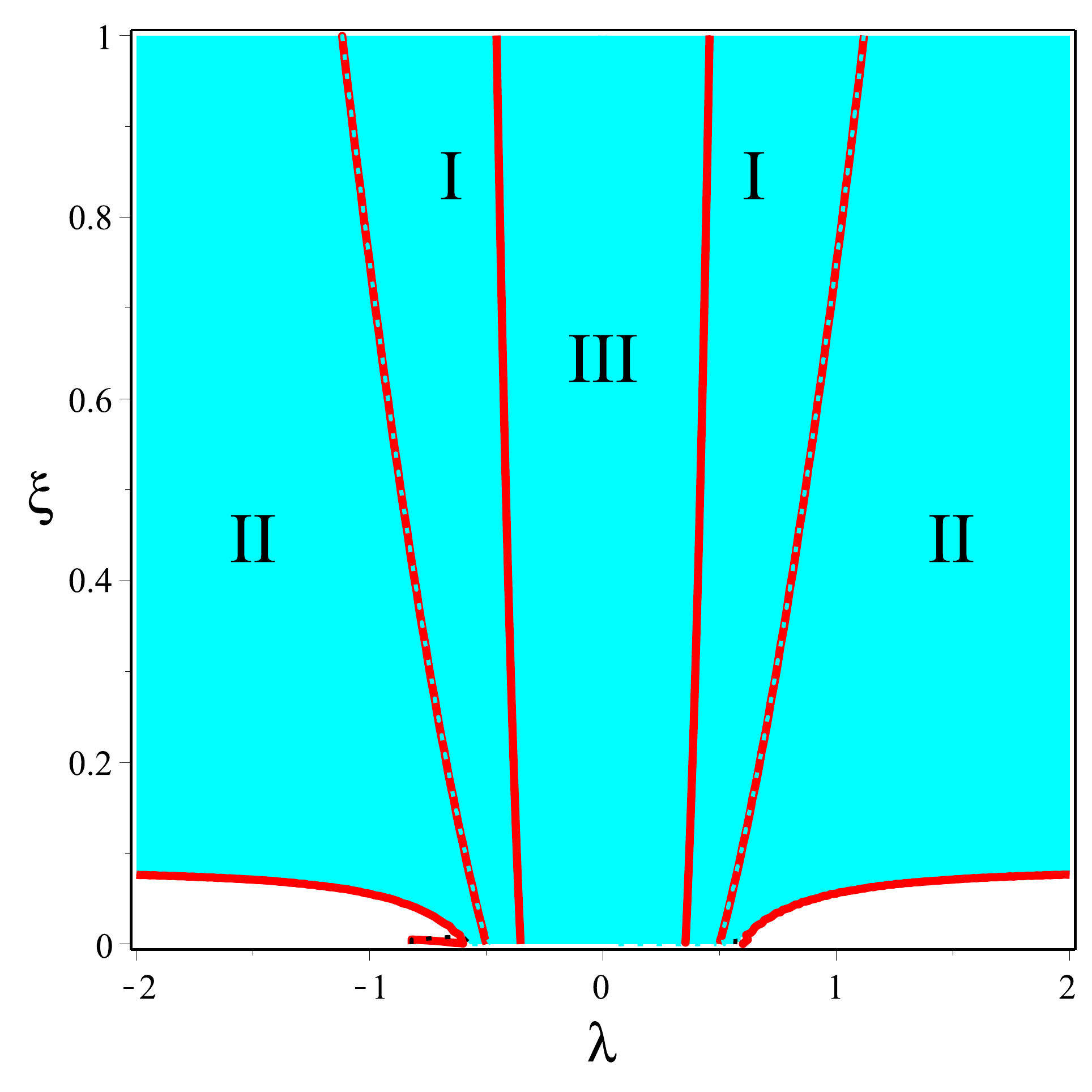}\label{existA3A41by3}}
\caption{Existence of points~$A_3$ and $A_4$ in the parameter space of the $V=V_0 \sinh^{-\alpha}(\lambda\phi)$ potential (in this figure $\alpha=4$ is assumed).
Panel (a) corresponds to the case $w=0$ and panel (b) corresponds to the case $w=\frac{1}{3}$. In the $w=0$ case, regions II and IV
corresponds to the existence region of $A_3$, while in the $w=\frac{1}{3}$ case point~$A_3$ does not exists at all. In both panels, regions I, II and III corresponds to the regions of existence of point $A_4$; region III denotes values of $\lambda$ and $\xi$ for which point~$A_4$ describes an accelerated solution, whereas its stability is attained only in regions I and III.  Here distinct numbered regions are separated by red solid curves.}
\label{A3A4}
\end{figure}
\begin{figure}
\centering
\subfigure[]{%
\includegraphics[width=5cm,height=4cm]{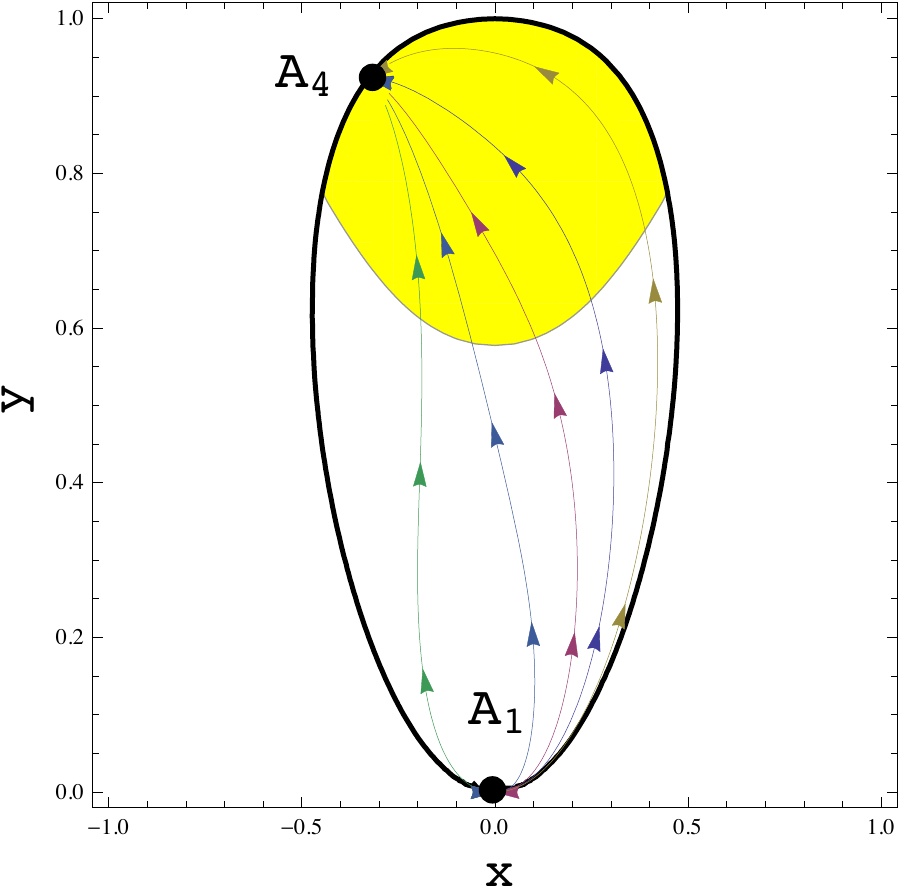}\label{contour_plot_1_1_A4}}
\qquad
\subfigure[]{%
\includegraphics[width=5cm,height=4cm]{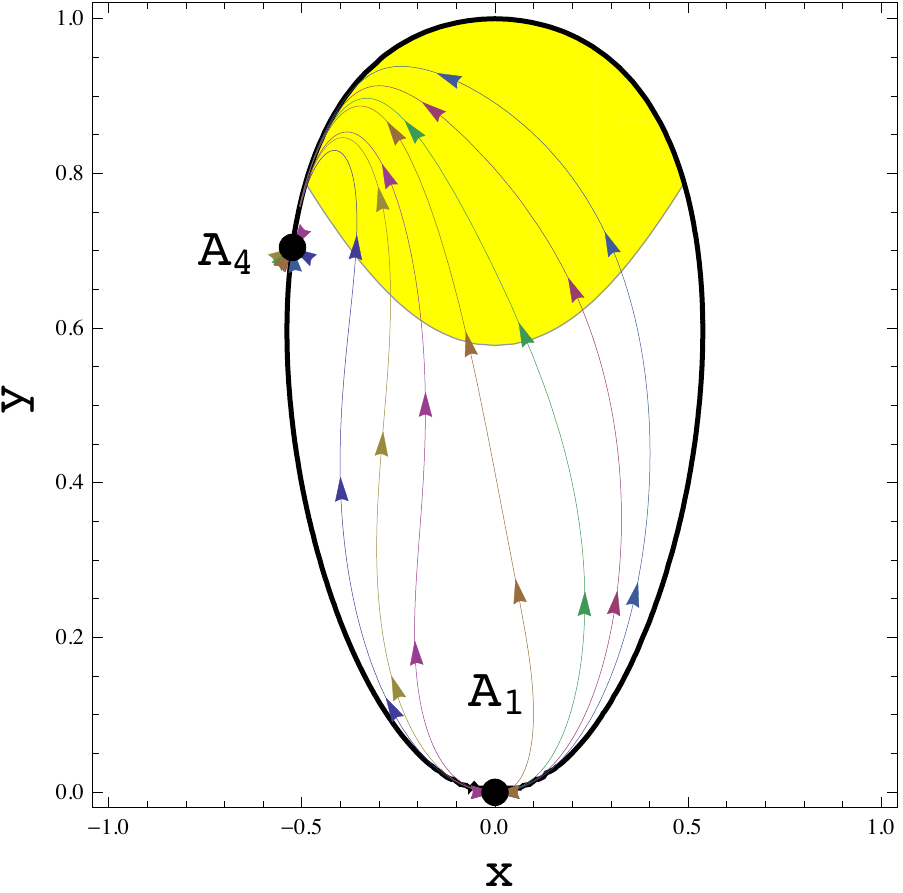}\label{contour_plot_2_1by2_A4}}
\qquad
\subfigure[]{%
\includegraphics[width=5cm,height=4cm]{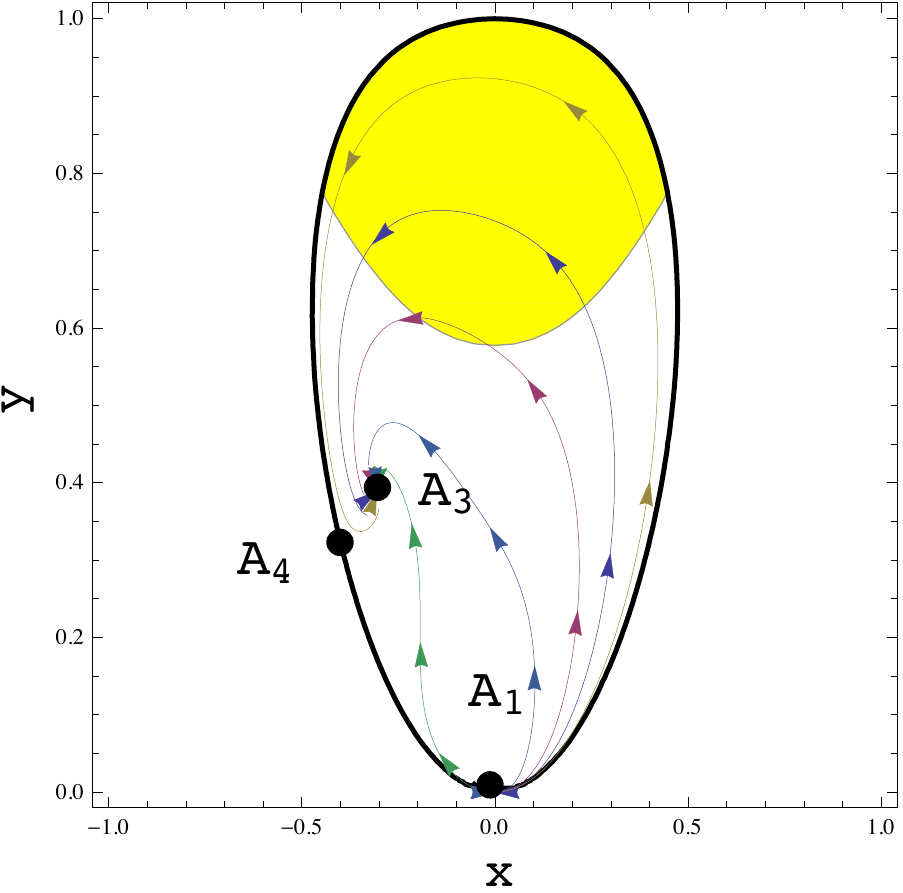}\label{contour_plot_4_1_A4.eps}}
\caption{
Projection on the plane $(x,y,s_*)$ of phase space trajectories of the system (\ref{eq23})-(\ref{eq25}) with $V=V_0 \sinh^{-\alpha}(\lambda\phi)$ and $w=0$. The yellow region represents the accelerated region of the phase space. The parameters have been chosen as follows: (a)~$\alpha=4$, $\lambda=-0.25$, $\xi=1$. ~(b)~$\alpha=8$, $\lambda=-0.25$, $\xi=0.5$.~(c)~$\alpha=16$, $\lambda=-0.25$, $\xi=1$.}
\label{spA4}
\end{figure}

For the potential \eqref{pot1} we have
\begin{equation}\label{eq28}
g(s)=\frac{1}{\alpha}-\frac{\alpha \lambda^2}{s^2} \,,
\end{equation}
so that
\begin{equation}\label{eq29}
s_*=\pm \alpha\lambda \,,~~~~~~~~ dg(s_*)=\frac{2\alpha\lambda^2}{s_*^3} \,.
\end{equation}
All critical points listed in Table~\ref{Tab1} are thus present in this case, with two copies of the points $A_3$ and $A_4$ for the two solutions $s_* = \pm \alpha\lambda$.

The properties of point $A_1$ do not depend on the scalar field potential, while the stability of point $A_2$ can now be fully investigated.
In this case in fact we can numerically plot projections on the $x,~y,~s$ axes of phase space trajectories as they approach point $A_2$ and check that for $N\rightarrow \infty$ they all attain the coordinates of point $A_2$. We observe from Figs.~\ref{fig1}-\ref{fig3} that projections on
$x,~y,~s$ of different trajectories approaching point $A_2$ do
indeed confirm that such point is stable as $N\rightarrow\infty$
(at least for $\xi = 1$ and $w=0$).
Point $A_2$ might thus represent the late time attractor of the phase space, where an accelerated scalar field dominated
solution ($\Omega_{\phi}=1,~w_{\rm eff}=-1$) takes place.
Moreover we have checked numerically that as long as $\alpha$ remains negative, for non-zero values of $w$ and different values of $\xi$ this situation does not change.
If instead $\alpha>0$ then point~$A_2$ becomes a saddle and the late time attractor is either point~$A_3$ or point~$A_4$.

According to the analysis above the scaling solution $A_3$ is a late time attractor if
\begin{equation}
    \frac{dg(s_*)}{s_*} = \frac{2}{\alpha^3 \lambda^2} > 0 \,,
\end{equation}
which translates into the condition $\alpha>0$. If instead
$\alpha<0$ then point $A_3$ is a saddle point, and the late time attractor is point~$A_2$.
Due to the complicated expressions of critical points $A_3$ and $A_4$,
we will only focus on the case of dust ($w=0$) and radiation ($w=\frac{1}{3}$) for their analysis.
The regions of existence of points $A_3$ and $A_4$ are shown in Fig.~\ref{A3A4}.
Point $A_3$ corresponds to an un-accelerated scaling solution with $w_{\rm eff}=w$, although it does not exist for any value of $(\lambda, \xi)$ when $w=\frac{1}{3}$ and $\alpha=4$, confirming the result we found without specifying a scalar field potential.
To give an example, by numerically choosing $\alpha=4,~\lambda=-\frac{1}{4},~\xi=1$, we obtain the eigenvalues $E_1=-2.2,~E_2=-2.59,~E_3=-0.4$ and the effective EoS $w_{\rm eff}=-0.73$.
This implies that point $A_4$ is a late time accelerated attractor for this choice of parameters (cf.~Fig.~\ref{A3A4}).
On the other hand if we take $\alpha=4,~\lambda=1,~\xi=1$, we obtain $E_1=0.9,~E_2=-1.04,~E_3=-1.95$ and $w_{\rm eff}=0.3$, which shows
that point $A_4$ is a saddle point (cf.~again Fig.~\ref{A3A4}).
In this latter case $A_3$ is the late time attractor.

Finally we plot projections of phase space trajectories of the system (\ref{eq23})-(\ref{eq25}) on the $s=s_*$ plane in Fig.~\ref{spA4}.
The dynamics on this plane is similar to the one appearing in the exponential potential case (cf.~\cite{Nicola}), although here we have a further dimension which renders the situation more complicated.
Nevertheless, by numerically plotting trajectories in the 3D phase space, we have checked that the $s=s_*$ plane is indeed an attractor for nearby trajectories.
Hence the dynamics on the $s=s_*$ plane can be taken as a description of the late time evolution of the universe in the case $\alpha>0$.
On such plane, one can realize that all trajectories start from the matter dominated point $A_1$, and eventually end either in point $A_4$ or, if it exists,
 in the scaling solution $A_3$.
In the first case, depending on the choice of the model parameters,
 we can find a late-time, accelerated, scalar field dominated solution, and thus a dynamical description of the dark matter to dark energy transition.

\begin{figure}[t]
\centering
\includegraphics[width=6cm,height=4cm]{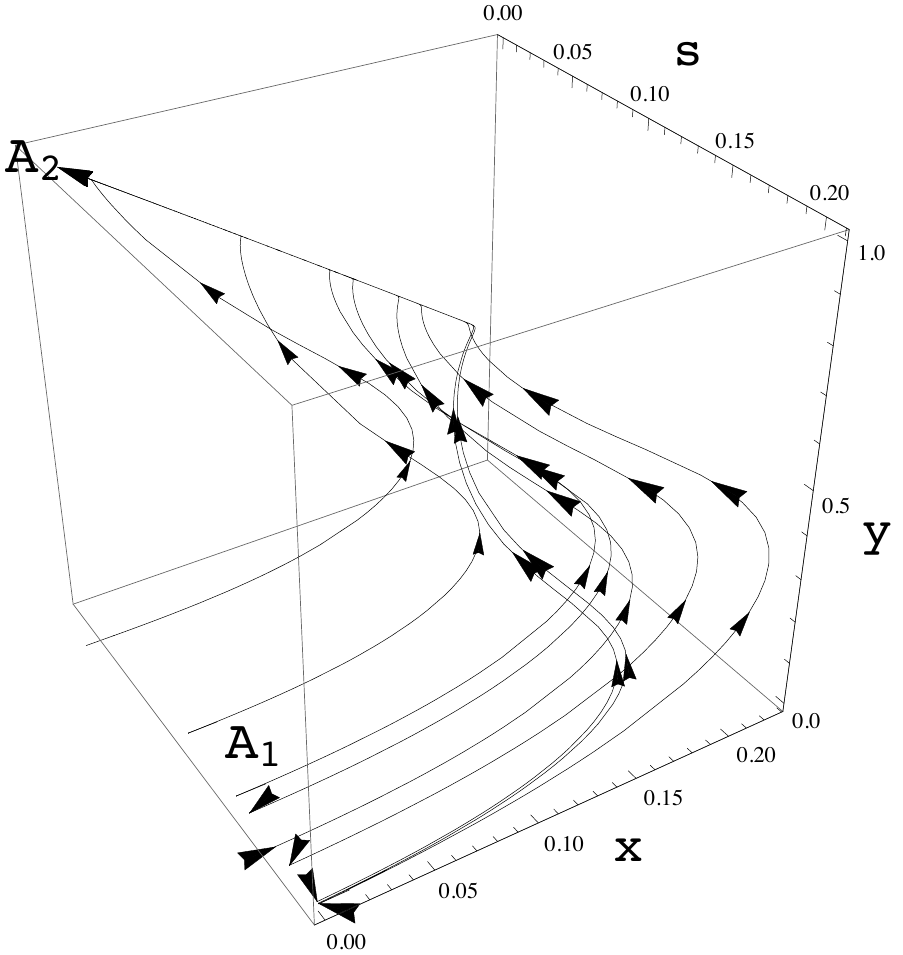}
\caption{Phase space of the system (\ref{eq23})-(\ref{eq25}) with the potential \eqref{eq:pot_pl}. All trajectories start from the matter dominated point $A_1~(0,0,s)$ (actually a critical line) and are attracted, along its centre manifold, towards point $A_2~(0,1,0)$, which describes an accelerating de Sitter solution. Here we have taken $w=0,~\xi=1,~n=4$.}
\label{sqkptp}
\end{figure}

\subsection{Example 2: $V=\frac{M^{4+n}}{\phi^n}$}
\label{subsec:square_2}

The second example that we consider is the potential
\begin{equation}
    V=\frac{M^{4+n}}{\phi^n} \,,
    \label{eq:pot_pl}
\end{equation}
where $M$ is a mass scale and $n$ a dimensionless parameter.
This potential can lead to tracking behaviour \cite{Zlatev}.
In this case we have simply
\begin{equation}\label{eq30}
g(s)=\frac{1}{n} \,,
\end{equation}
implying that there does not exist any solution of $g(s)=0$.
Therefore for this model $s_*$ does not exist at all, and we have only two critical points: a matter dominated point $A_1$ (saddle)
and an accelerated scalar field dominated point $A_2$ which constitutes the late time attractor (cf.~Fig.~\ref{sqkptp}).
The dynamics in the phase space of this cosmological model is thus
quite simple. All physically interesting trajectories pass near point $A_1$ and eventually reach the
late time attractor $A_2$, following its centre manifold, as shown in Fig.~\ref{sqkptp}.
In fact the trajectories that connects point $A_1$ (formally a critical line \cite{TamaniniPhDthesis}) with point $A_2$
represent a viable realisation of the observed universe, where a
transition from dark matter to dark energy domination can take
place.
Note that this happens for a wide range of initial conditions, implying that fine tuning issues might be somehow less severe for this particular model.


\section{Square Root Kinetic Corrections}
\label{sec:sqr_r_kc}

In this section, we consider square root kinetic corrections to the canonical scalar field Lagrangian, in other words we assume $n=1/2$ in \eqref{eq:001}.
The scalar field Lagrangian is thus given by
\begin{equation}\label{eq31}
\mathcal{L}_{\phi}=X-V+\xi \sqrt{XV} \,,
\end{equation}
where the parameter $\xi$ can take any real value.
Using again the dimensionless variables \eqref{eq13}, the Friedmann equation (\ref{eq14}) can be written as
\begin{equation}\label{eq32}
1=x^2+y^2+z^2 \,,
\end{equation}
which is nothing but the constraint one would get in the canonical case.
Again we can use \eqref{eq32} to substitute $z^2$ in all equations that follows, effectively reducing the dimension of the system from four to three.
In terms of the dimensionless variables (\ref{eq13}) the following relevant cosmological parameters viz., the relative scalar field energy density parameter, the relative energy density parameter of matter, the EoS parameter of scalar field, the effective EoS parameter and deceleration parameter, are respectively given by
\begin{eqnarray}
\Omega_{\phi}&=&x^2+y^2\label{eq33}\,,\\
\Omega_{m}&=&1-x^2-y^2\label{eq34}\,,\\
w_{\phi}&=&\frac{x^2-y^2+\xi x y}{x^2+y^2+\xi(2n-1)x y}\label{eq35}\,,\\
w_{\rm eff}&=&x^2-y^2+w(1-x^2-y^2)+\sqrt{2}\xi x y\label{eq36}\,,\\
q&=&-1-\frac{\dot{H}}{H^2}=-1-\frac{3}{2}\left[(w-1)x^2+(w+1)(y^2-1)-\sqrt{2}\xi x y\right]\label{eq36a}\,.
\end{eqnarray}
Moreover the adiabatic speed of sound is given by
\begin{equation}\label{eq37}
C_s^2=1+\frac{\xi y}{2 x} \,.
\end{equation}
We can immediately notice that whenever $X=0$, i.e.~$x=0$, the adiabatic speed of sound diverges, meaning that adiabatic perturbations travel at infinite velocity.
Furthermore whenever $\xi x<0$, we obtain $C_s^2<0$ in some region of the phase space, implying instability at the classical level.
These results imply that this model is not viable at the physical level and cannot constitute a sensible description of Nature.
Nevertheless in what follows we will ignore such problems and analyse the dynamical systems obtained from this cosmological model, for both the sake of mathematics and the possibility that some model will produce the same background dynamics while being physically consistent, as indeed happens using an exponential potential \cite{Boehmer:2015sha}.
For numerical solutions, we again estimate the initial conditions in such a way that they match with the present observational data,
corresponding to points in phase space where
\begin{equation}\label{37a}
x_0=\pm 0.52,~~y_0=\pm 0.67 \,.
\end{equation}

\begin{center}
\begin{table}[t]
\caption{Critical points and corresponding cosmological parameters
of the system (\ref{eq38})-(\ref{eq40}). Here we define:
$\eta=\sqrt{\xi^2-2w^2+2}$,  $\chi=\sqrt{3\xi^2+6-s_*^2}$,
$E=\frac{1}{3}\frac{2s_*^2-3(\xi^2+2)- s_* \xi
\sqrt{2}\chi}{(\xi^2+2)}$, $\Omega_{\phi_8}=\frac{3}{2}\,\frac
{({\xi}^{2}+2\,w+2)\pm\xi\,\eta}{{s _{*}}^{2}}$. }.
\begin{center}
\begin{tabular}{c c c c  c c c c}
\hline\hline
Point  &  $~~~x~~~$   &$~~~y~~~$     &$~~~s~~~$    & ~~~Existence    ~~~&$~~~\Omega_\phi~~~$    &$~~~\omega_{\rm eff}~~~$   &Acceleration \\ \hline\\
$B_1$&  $0$   &$0$    &$s$      &Always        & $0$             & $w$            &No\\ [1.5ex]
$B_2$&  $1$    &$0$   &$0$    &Always   & $1$             & $1$            &No\\ [1.5ex]
$B_3$&  $-1$    &$0$   &$0$    &Always   & $1$             & $1$            &No \\ [1.5ex]
$B_4$&  $-\xi\,\sqrt{\frac{1}{\xi^2+2}}$    &$\sqrt\frac{2}{\xi^2+2}$   &$0$    &Always   & $1$             & $-1$            &Always \\ [1.5ex]
$B_5$&  $0$    &$1$   &$\sqrt{3}\xi$    &Always   & $1$             & $-1$            &Always\\ [1.5ex]
$B_6$ & $1$ & $0$ & $s_*$ & Always& $1$& $1$ & No\\[1.5ex]
$B_7$ & $-1$ & $0$ & $s_*$ & Always& $1$& $1$ & No\\[1.5ex]
$B_8$& $\sqrt{\frac{3}{2}}\frac{(w+1)}{s_*}$   &$\frac{\sqrt{3}}{2
s_*}\left(\xi\pm\eta\right)$  &$s_*$   & Fig. \ref{B8B9s} &
$\Omega_{\phi_8}$    & $w$      &No\\ [1.5ex] $B_9$&
$\frac{\sqrt{2}s_*-\xi\chi}{\sqrt{3}(\xi^2+2)} $    &$\frac{s_*
\xi+\sqrt{2}\chi}{\sqrt{3}(\xi^2+2)}$   &$s_*$    &Fig.
\ref{B8B9s}  &$1$          &      $E$       &Fig.\ref{B8B9s}\\
[1.5ex]
\hline\hline\\[1.5ex]
\end{tabular}\label{Tab3}
\end{center}
\end{table}
\end{center}

Using the dimensionless variables (\ref{eq13}), the cosmological equations of this model can be written as the following autonomous system of equation
\begin{eqnarray}
x'&=&\frac{1}{2}\left[-3(w-1)x^3-3x\left((w+1)y^2-w+1\right)+3\sqrt{2}\xi\,x^2y+\sqrt{2}y\left(\sqrt{3}s\,y-3\xi\right)\right]\label{eq38}\,,\\
y'&=&-\frac{1}{2}y\left[3(w-1)x^2+3(w+1)(y^2-1)+x\left(\sqrt{6}s-3\sqrt{2}\xi\,y\right)\right]\label{eq39}\,,\\
s'&=&-\sqrt{6}\,x\,s^2\,g(s)\,.\label{eq40}
\end{eqnarray}
Note that the system (\ref{eq38})-(\ref{eq40}) is invariant under the simultaneous transformation $x\rightarrow-x$, $\xi\rightarrow-\xi$ and $s\rightarrow-s$, meaning that the dynamics in the $x s<0$ region mirrors the one in the $x s>0$ region for opposite values of $\xi$.
Moreover it can also be seen that the system (\ref{eq38})-(\ref{eq40}) is invariant under the transformation $y\rightarrow-y$, $\xi\rightarrow-\xi$, implying that the negative $y$ half of the phase space present the same dynamics of the positive half for opposite values of $\xi$.
Again from the physical requirement $z^2\geq 0$, we obtain the constraint
\begin{equation}\label{eq41}
0\leq x^2+y^2\leq 1 \,,
\end{equation}
which tells us that the phase space of the system (\ref{eq38})-(\ref{eq40}) is given by
\begin{equation}\label{eq42}
\Psi=\left\lbrace (x,y)\in \mathbb{R}^2: 0\leq x^2+y^2\leq 1\right\rbrace \times \left\lbrace s \in \mathbb{R}\right\rbrace \,.
\end{equation}

\begin{center}
\begin{table}[t]
\caption{Eigenvalues of the critical points listed in Table \ref{Tab3}.}.
\begin{center}
\begin{tabular}{c c c c c}
\hline\hline
Point  &  $E_1$   &$E_2$     &$E_3$  &Stability  \\ \hline\\
$B_1$&  $\frac{3}{2}(w-1)$   &$\frac{3}{2}(w+1)$    &$0$& saddle     \\
[1.5ex] $B_2$&   $3(1-w)$    &$3$    &$0$ & unstable/saddle   \\
[1.5ex] $B_3$&  $3(1-w)$    &$3$  &$0$  &  unstable/saddle  \\
[1.5ex]
$B_4$&  $-3(1+w)$    &$-3$   &$0$& stable/saddle   \\ [1.5ex]
$B_5$&  $-\frac{3}{2}\left(1-\sqrt{1-4 \xi^2\,g(\sqrt{3}\xi)}\right)$    &$-\frac{3}{2}\left(1+\sqrt{1-4 \xi^2\,g(\sqrt{3}\xi)}\right)$   &$-3(w+1)$    & stable/saddle \\ [1.5ex]
$B_6$ & $3(1-w)$ & $3-\frac{\sqrt{6}}{2}s_*$&$-\sqrt{6} s_* dg(s_*)$& unstable/saddle\\[1.5ex]
$B_7$ & $3(1-w)$ & $3+\frac{\sqrt{6}}{2}s_*$&$-\sqrt{6} s_* dg(s_*)$& unstable/saddle\\ [1.5ex]
$B_8$  &$\tau_+$   &$\tau_-$  &  $-3(1+w)s_* dg(s_*)$    & stable/saddle  \\ [1.5ex]
$B_9$  &$\zeta_+$   &$\zeta_-$    &  $\frac{-\sqrt{2}\left(\sqrt{2}s_*-\xi\chi\right)s_*^2 dg(s_*)}{\xi^2+2} $  & stable/saddle  \\ [1.5ex]
\hline\hline\\[1ex]
\end{tabular}\label{Tab4}
\end{center}

$\tau_{\pm}=-\frac{3}{4}\left[(1-w)\,\pm \frac{1}{s_*}\left(4{s_{*}}^{2}\xi\,\eta+12 {w}^{2}\xi\,\eta -12{\xi}^{3}\,\eta-12 w\xi\,\eta+9\,{s_{*}}^{2}{w}^{2}-4\,{s_{*}}^{2}{\xi}^{2}-24\,{w}^{2}{\xi}^{2}\right.\right.$\\$\left.\left.+12\,{\xi}^{4}
-24 \xi\,\eta-2\,w{s_{*}}^{2}-24\,{w}^{3}+12\,w{\xi}^{2}-7\,{s_{*}}^{2}-24\,{w}^{2}+36\,{\xi}^{2}+24\,w+24\right)^\frac{1}{2}\right]$\\[1.5ex]

$\zeta_{\pm}=\frac{1}{{\xi}^{2}+2}\left[-\frac{3}{4}\sqrt {2}\,\chi s_{*}\,\xi-\frac{3}{2}w{\xi}^{2}+\frac{3}{2}\,{s_{*}}^{2}-3\,{\xi}^{2}-3\,w-6\pm\frac{1}{2}\left(6\,\sqrt {2}\,\chi s_{*}\,w{\xi}^{3}-2\,\sqrt {2}\chi {s_{*}}^{3}\xi- {s_{*}}^{4}{\xi}^{2}\right.\right.$\\$\left.\left.+3\,{s_{*}}^{2}{\xi}^{4}+18\,{w}^{2}{\xi}^{4}+12\,\sqrt {2}\,\chi s_{*}\,w\xi-12\,{s_{*}}^{2}w{\xi}^{2}+2\,{s_{*}}^{4}+6\,{s_{*}}^{2}{\xi}^{2}+72\,{w}^{2}{\xi}^{2}-24\,w{s_{*}}^{2}+72\,{w}^{2}\right)^\frac{1}{2}\right]$\\[1.5ex]

\end{table}
\end{center}

\begin{figure}[t]
\centering
\subfigure[]{%
\includegraphics[width=8cm,height=6cm]{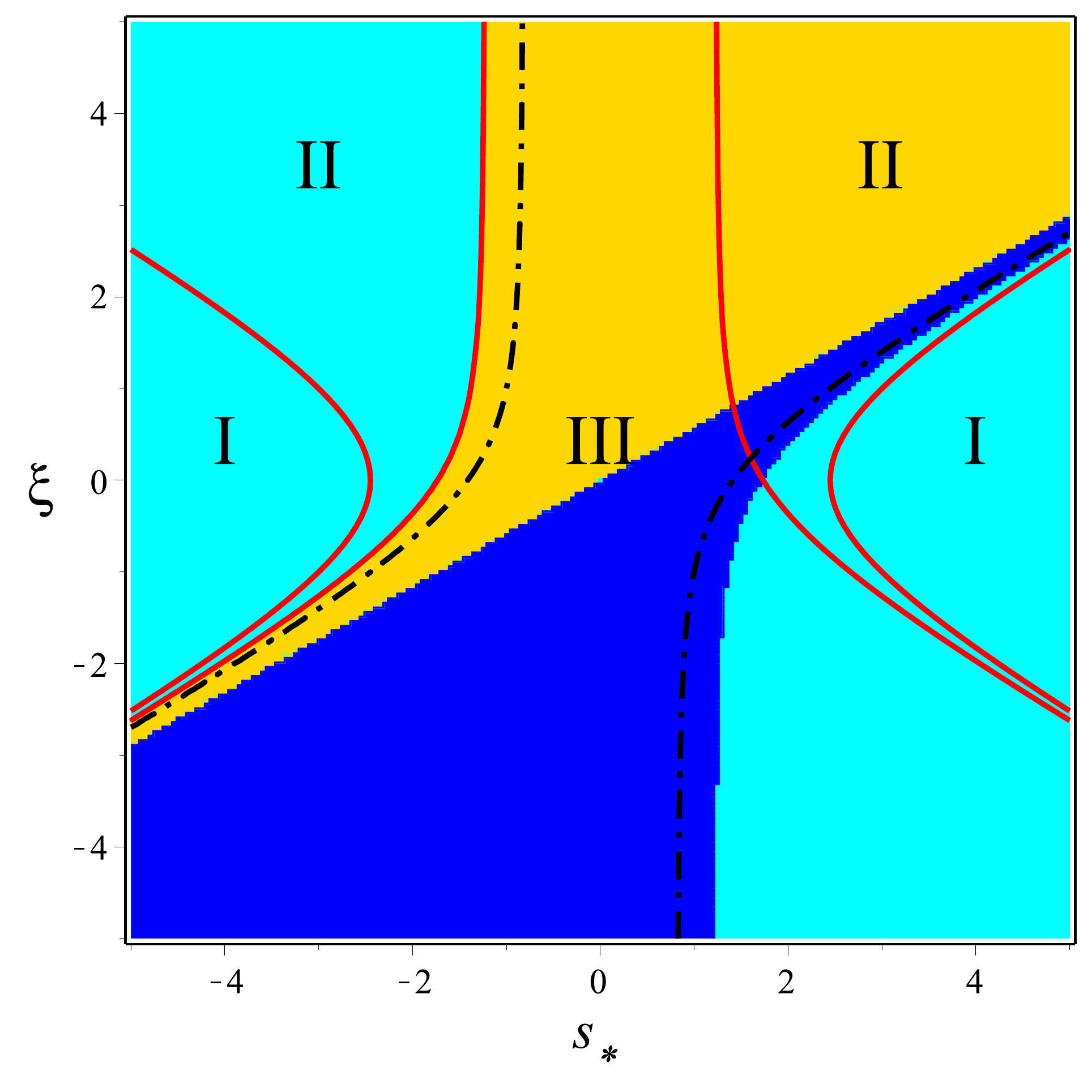}\label{B8B9zeros}}
\qquad
\subfigure[]{%
\includegraphics[width=8cm,height=6cm]{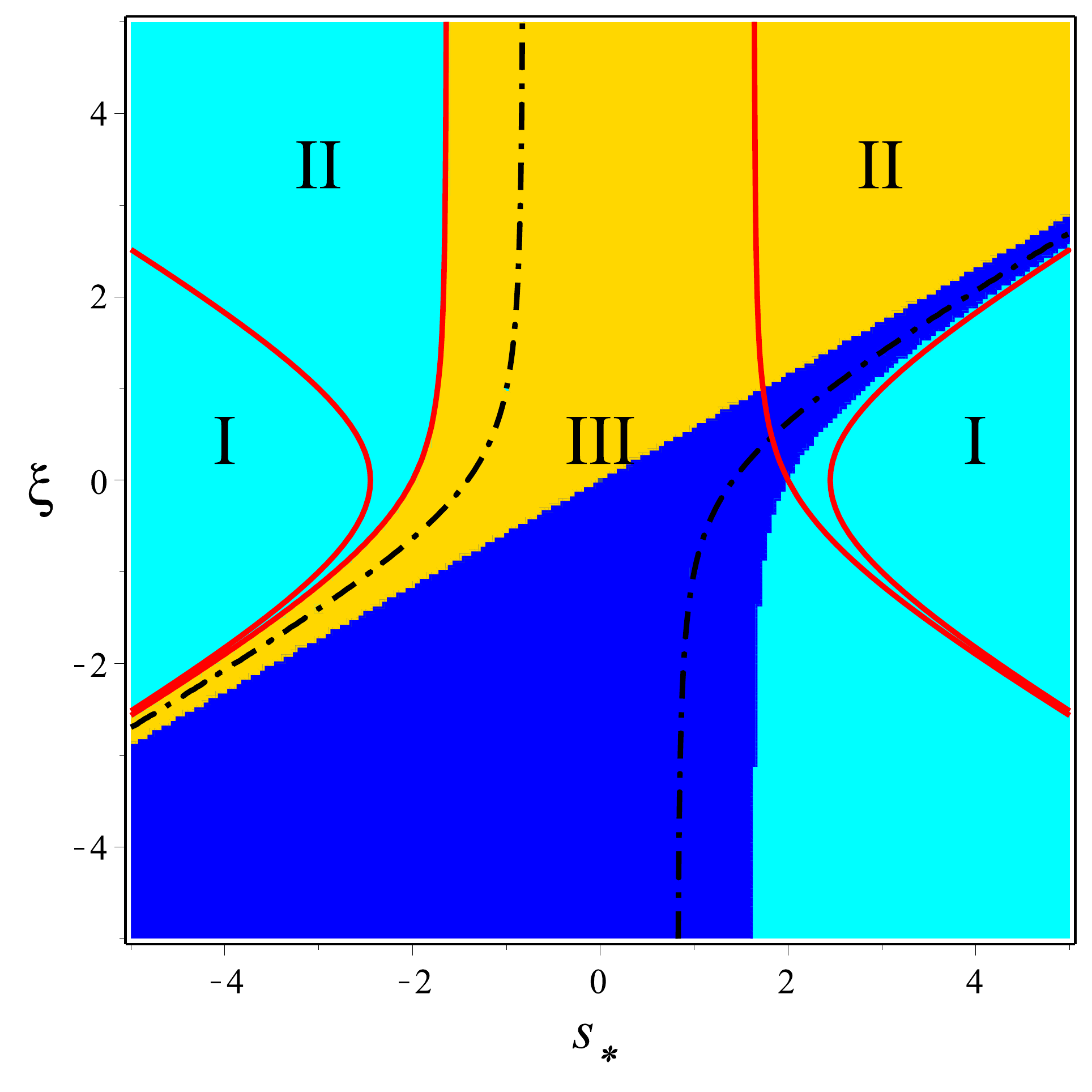}\label{B8B91by3s}}
\caption{Existence and stability in the parameter space of points $B_8$ and $B_9$ for the system (\ref{eq38})-(\ref{eq40}) . Panel (a) corresponds to the case $w=0$ and panel (b) corresponds to the case $w=\frac{1}{3}$. In both panels regions I and II are the existence regions of point $B_8$, regions II and III are the existence regions of point $B_9$. The yellow shaded region represents the region of stability of point $B_9$ for potentials where $dg(s_*)<0$, whereas the dark blue region represents the region of stability of point $B_9$ for potentials where $dg(s_*)>0$; the internal region bounded by dot-dashed curves represents the acceleration region of point $B_9$. Here distinct numbered regions are separated by red solid curves.}\label{B8B9s}
\end{figure}
\begin{figure}[t]
\centering
\subfigure[]{%
\includegraphics[width=5cm,height=3.75cm]{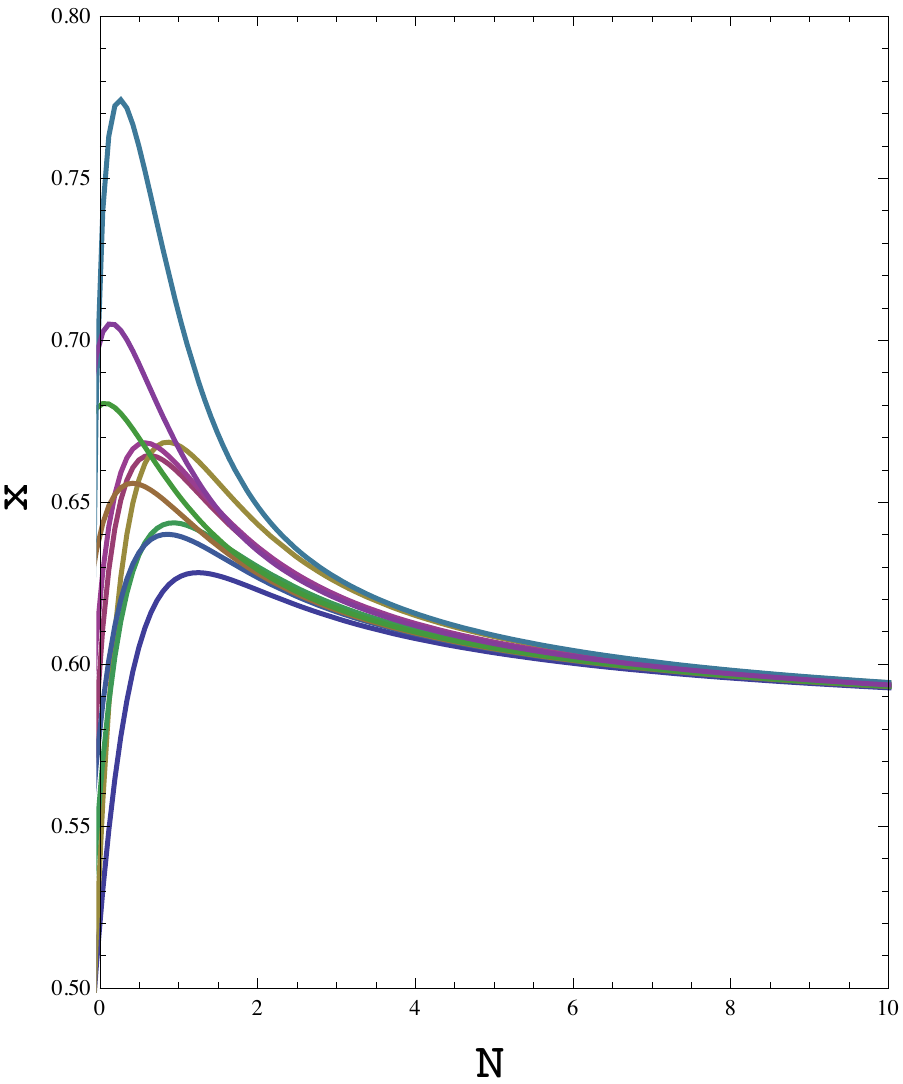}\label{fig4}}
\qquad
\subfigure[]{%
\includegraphics[width=5cm,height=3.75cm]{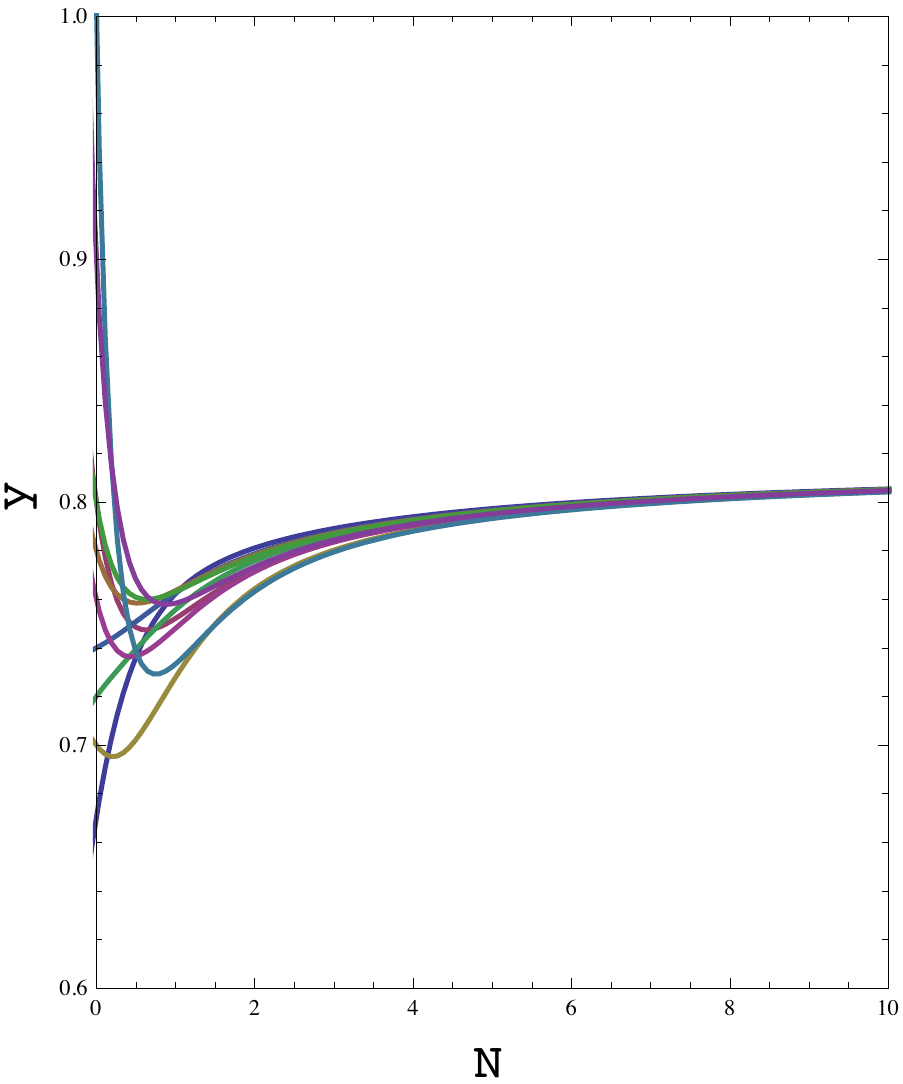}\label{fig5}}
\qquad
\subfigure[]{%
\includegraphics[width=5cm,height=3.75cm]{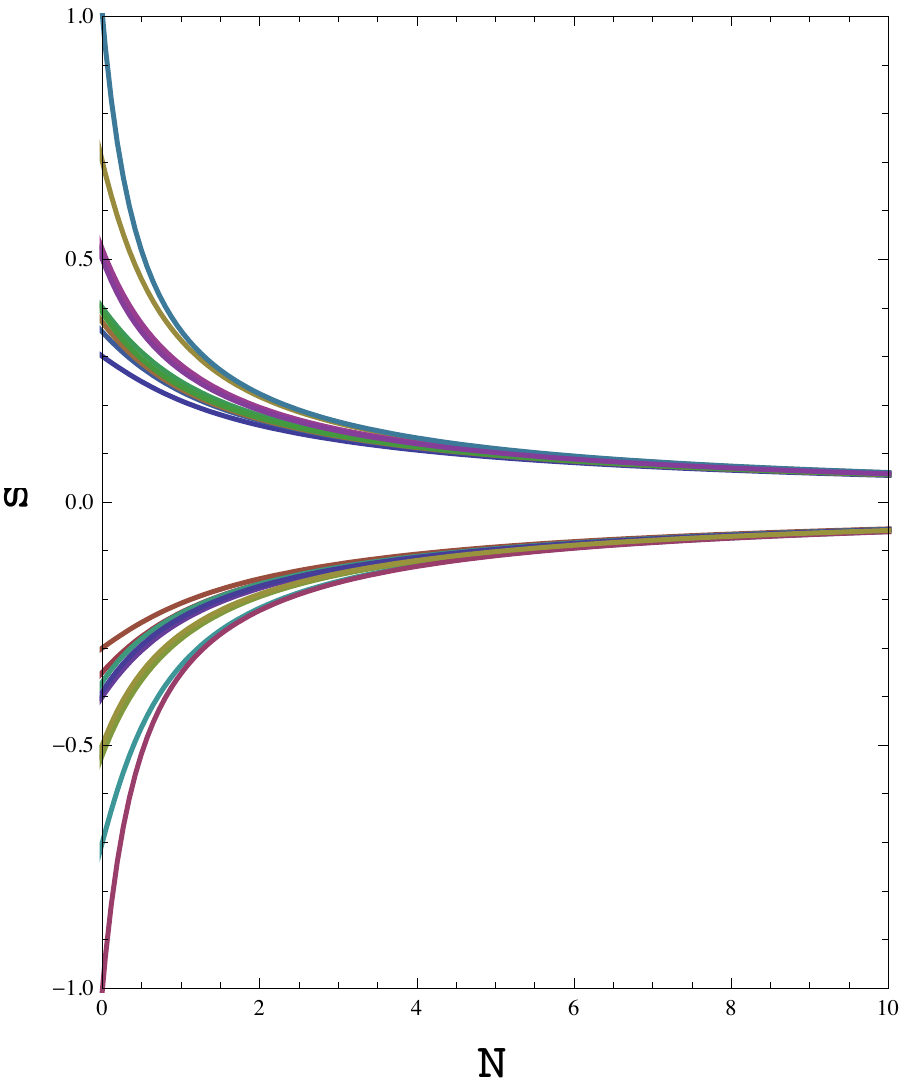}\label{fig6}}
\caption{(a).~Time evolution of $x$-projections of trajectories approaching point $B_4$; (b).~Time evolution of $y$-projections of trajectories approaching point $B_4$; (c).~Time evolution of $s$-projections of trajectories approaching point $B_4$. Here we have considered the potential \eqref{pot1} with $\xi=-1$, $\alpha=1$, $\lambda=0$.}
\end{figure}

\begin{figure}[t]
\centering
\subfigure[]{%
\includegraphics[width=8cm,height=6cm]{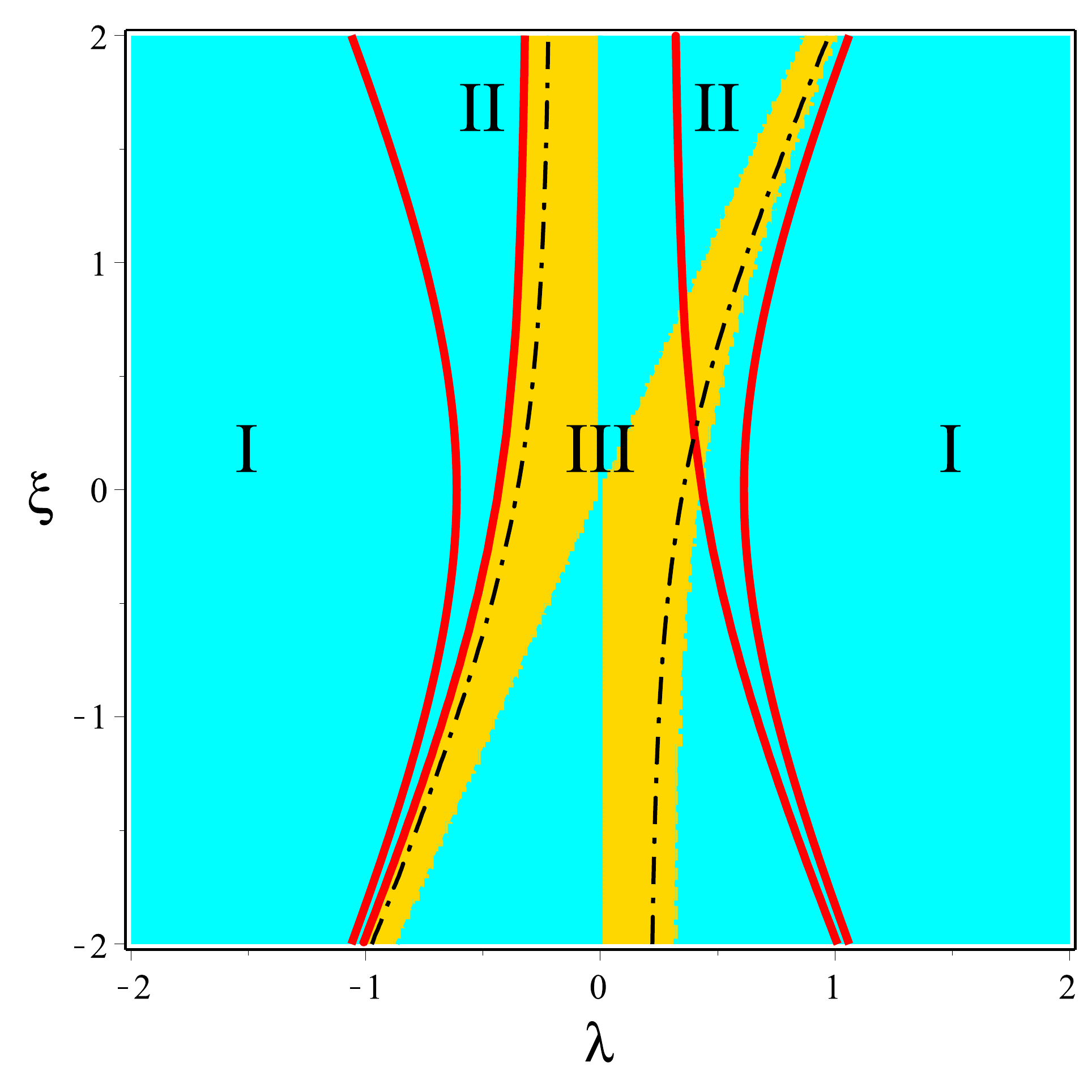}\label{B8B9zero}}
\qquad
\subfigure[]{%
\includegraphics[width=8cm,height=6cm]{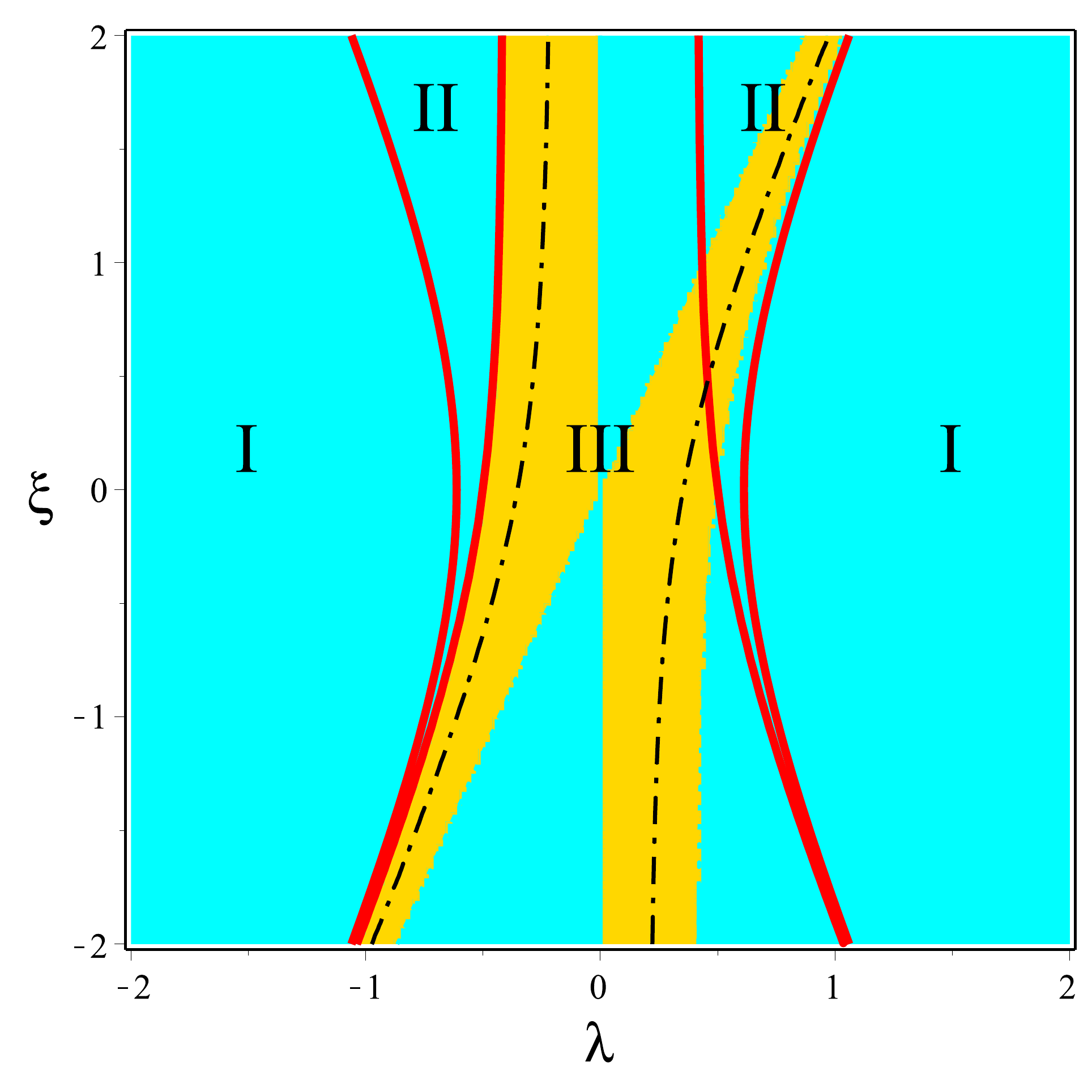}\label{B8B91by3}}
\caption{Existence and stability in the parameter space of points $B_8$ and $B_9$ for the system (\ref{eq38})-(\ref{eq40}) with potential $V=V_0 \sinh^{-\alpha}(\lambda\phi)$. Panel (a) corresponds to the case $w=0$ and panel (b) corresponds to the case $w=\frac{1}{3}$. In both panels regions I and II are the existence regions of point $B_8$, regions II and III are the existence regions of point $B_9$. The yellow shaded region represents the region of stability of point $B_9$, while the internal region bounded by dot-dashed curves represents the acceleration region of point $B_9$. Here we have taken $\alpha=4$.}\label{B8B9}
\end{figure}
\begin{figure}[t]
\centering
\subfigure[]{%
\includegraphics[width=5cm,height=4cm]{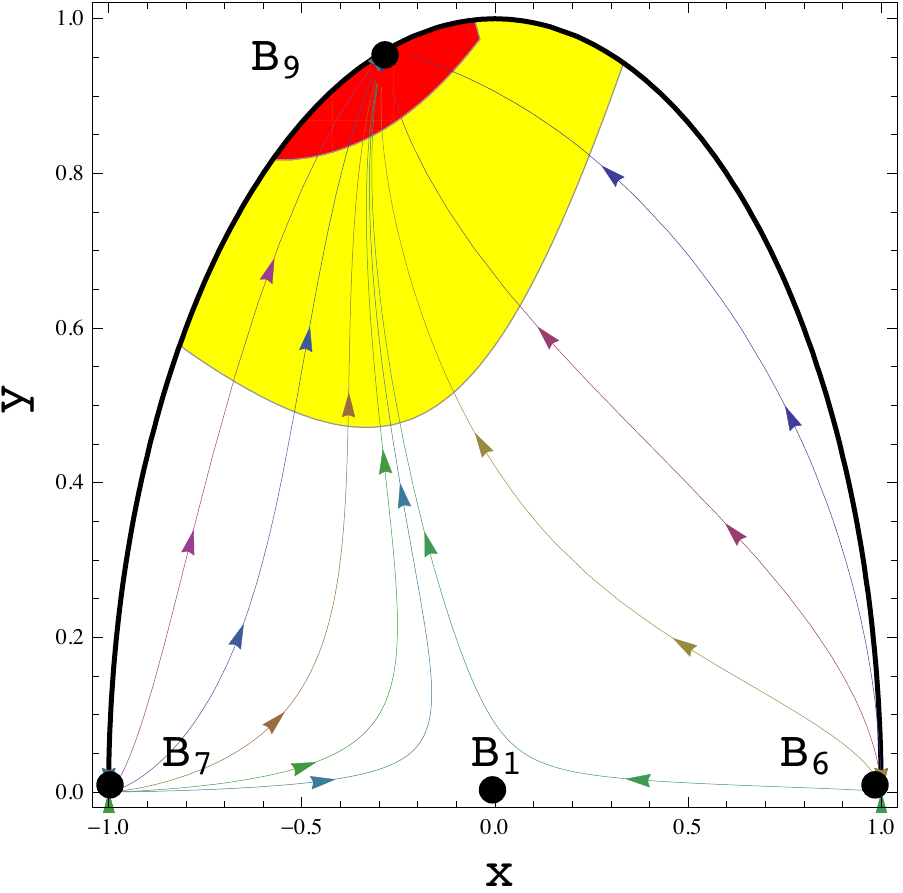}\label{contour_plot_B9_1_1}}
\qquad
\subfigure[]{%
\includegraphics[width=5cm,height=4cm]{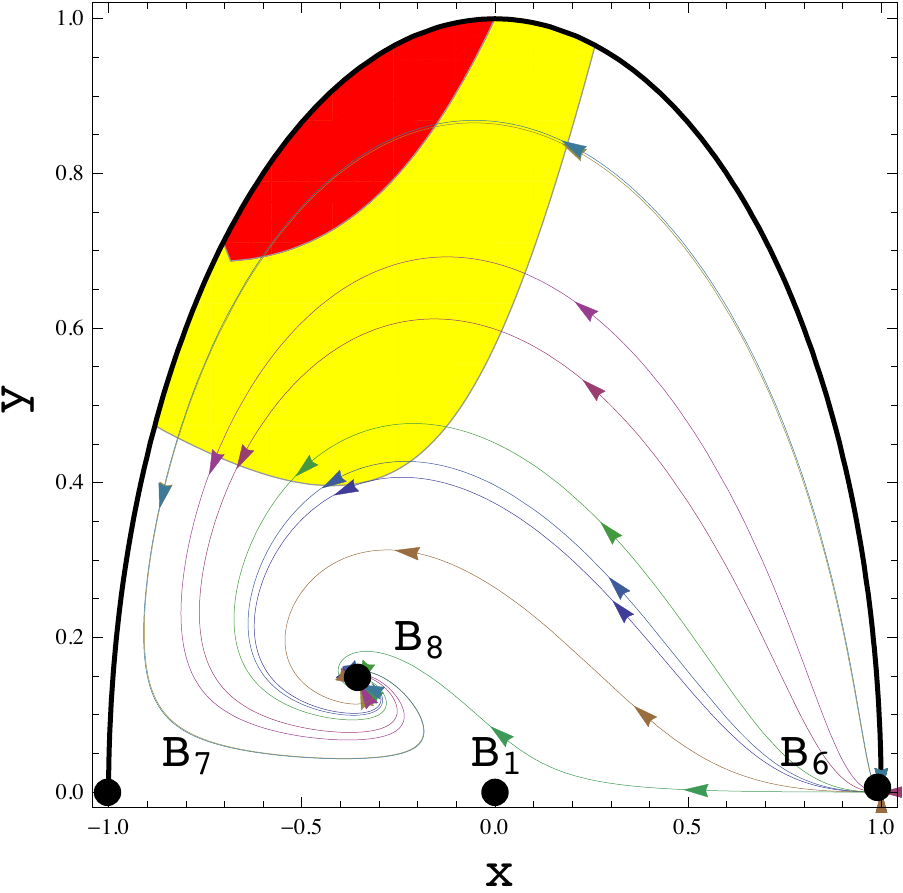}\label{contour_plot_B9_3_1}}
\qquad
\subfigure[]{%
\includegraphics[width=5cm,height=4cm]{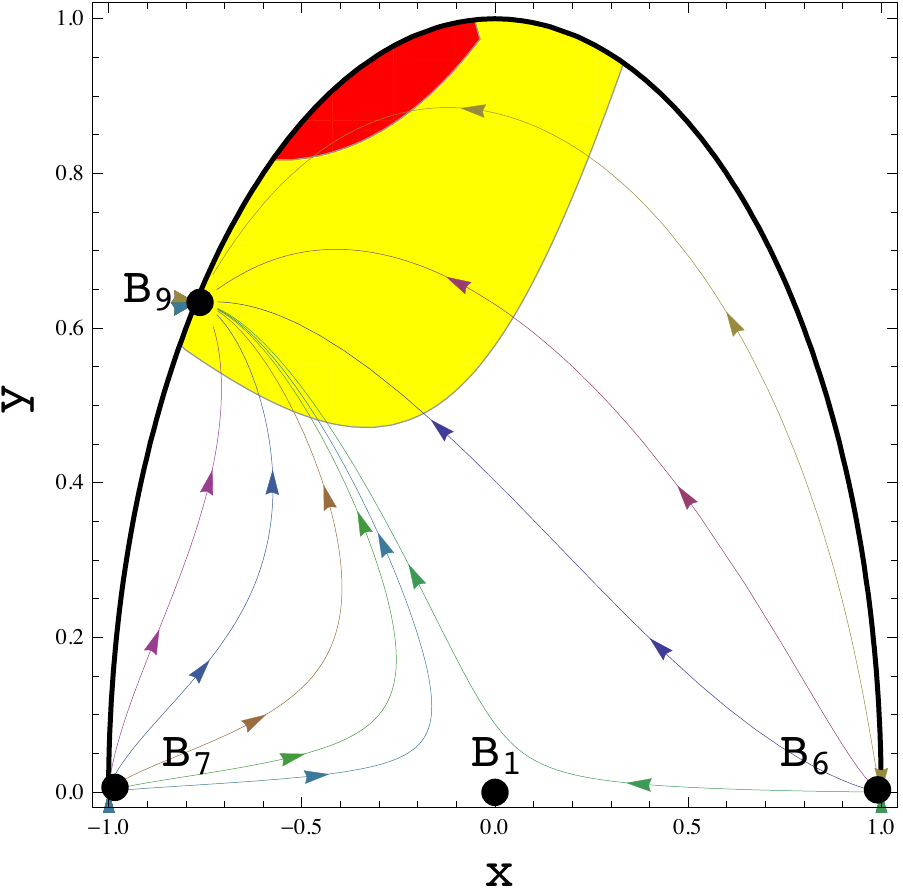}\label{contour_plot_B9}}
\qquad
\subfigure[]{%
\includegraphics[width=5cm,height=4cm]{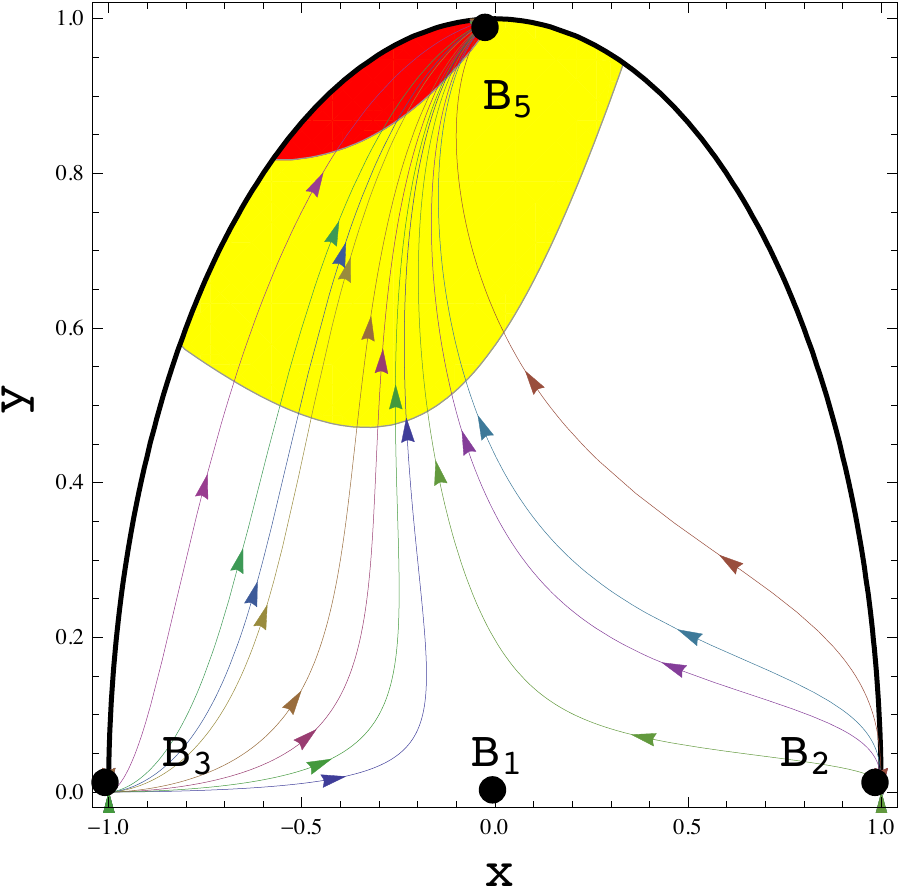}\label{contour_plot_B5}}
\qquad
\subfigure[]{%
\includegraphics[width=5cm,height=4cm]{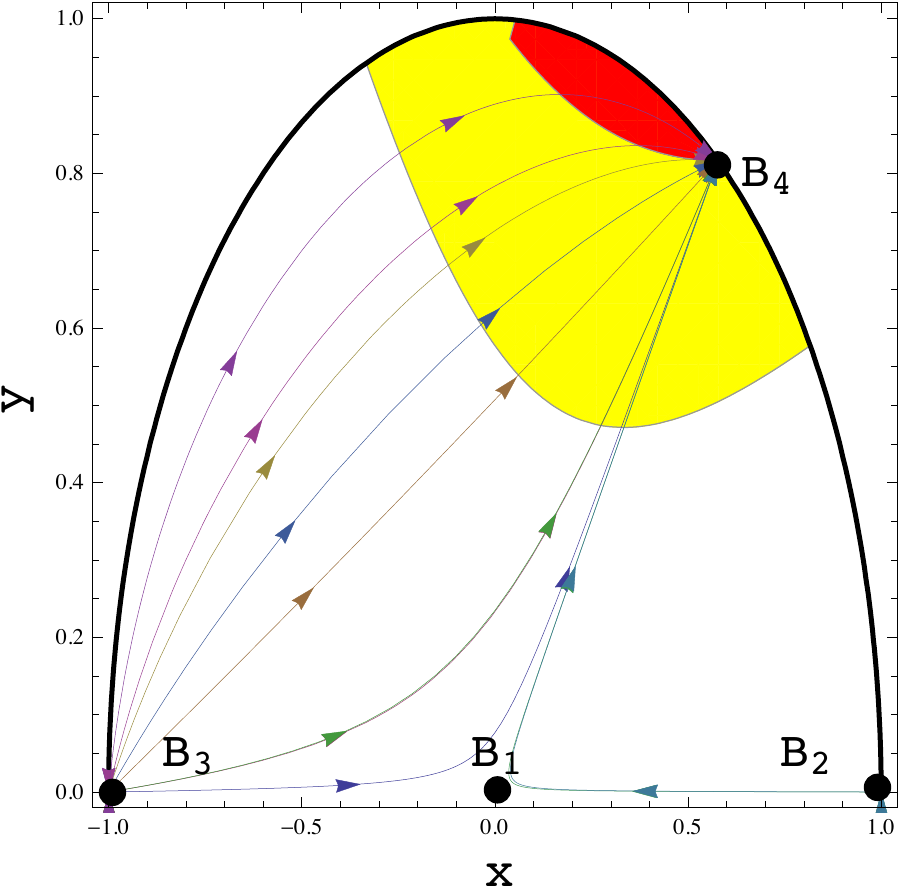}\label{contour_plot_B4}}
\caption{Projections of phase space trajectories of the system (\ref{eq38})-(\ref{eq40}) on a plane $(x,y,s_*)$ with $w=0$ and the potential \eqref{pot1}. The yellow region represents the accelerated region of the phase space. The red region represents the phantom accelerated region of the phase space. In these plots we have assumed the following values: (a)~$\alpha=-4$, $\lambda=0.25$, $\xi=1$; (b)~$\alpha=4$, $\lambda=-\frac{7}{8}$, $\xi=1.5$; (c)~$\alpha=4$, $\lambda=-0.2$, $\xi=1$; (d)~$\alpha=\sqrt{3}$, $\lambda=1$, $\xi=1$; (e)~$\alpha=1$, $\lambda=0$, $\xi=-1$. }
\label{spB9}
\end{figure}

The critical points of the system (\ref{eq38})-(\ref{eq40}) along with their relevant cosmological parameters are summarised in Table~\ref{Tab3}, while the eigenvalues of the corresponding Jacobian matrices are given in Table~\ref{Tab4}.
There are at least five critical points (points $B_1$--$B_5$), irrespectively of the form of the potential $V(\phi)$, while a number of further critical points (points $B_6$--$B_9$) will appear depending on how many solutions $s_*$ there will be of the equation $g(s)=0$.
Critical point $B_1$ exists for an arbitrary potentials, while for points $B_2,~B_3,~B_4$ the potential is effectively constant since its derivative with respect to $\phi$ vanishes.
For point $B_5$ to exist the scalar field potential must depend on the parameter $\xi$, whereas the existence of points $B_6-B_9$ depends on the concrete form of the potential through $s_*$. It is noted that critical point $B_9$ reduces to point $B_4$, when $s_*=0$.

Points~$B_1$--$B_4$ are all non-hyperbolic critical points.
Point $B_1$ is a saddle point corresponding to a matter dominated
solution. Point $B_2,~B_3$ behave instead as unstable nodes or saddles (depending on the stability of their centre manifolds) and
correspond to stiff fluid solutions ($w_{\rm eff} = 1$) dominated
by the kinetic part of the scalar field. Point~$B_4$ describes an accelerated scalar field
dominated solution.
Its stability properties can only be determined analytically once the dynamics along its centre manifold is analysed.
However, given a particular potential, it is easier to check the stability of point $B_4$ numerically, and we will thus leave this analysis to be completed in any particular case, as in the examples below.

Point $B_5$ corresponds to an
accelerated scalar field dominated solution. It is a late time
attractor if $g(\sqrt{3\xi})>0$, but it is a saddle for
$g(\sqrt{3}\xi)<0$. Critical points $B_6$ and $B_7$ correspond to stiff matter dominated solutions. Point $B_6$ is an unstable node
for $s_*<\sqrt{6}$ and $s_* dg(s_*)<0$ otherwise it is a saddle
point. Point $B_7$ is an unstable node for $s_*>-\sqrt{6}$ and
$s_* dg(s_*)<0$ otherwise it is a saddle point. Critical point
$B_8$ describes a decelerated scaling solution.
Its existence region in the $(\xi, s_*)$ parameter space is given in Fig.~\ref{B8B9s}.
For this point,
it can be checked that the eigenvalues $E_1,~E_2$ have always
negative real part within the region of existence of the critical
point. On the other hand $E_3$ is positive or negative depending
on whether $s_*dg(s_*)<0$ or $s_*dg(s_*)>0$. So point $B_8$ is
stable if $s_*dg(s_*)>0$ and it is a saddle point if
$s_*dg(s_*)<0$.
Finally point $B_9$ corresponds to a scalar field dominated solution.
Again, due to the complicated expressions of its eigenvalues, we can only check the existence and stability of point~$B_9$ numerically by choosing different values of $\xi$ and $s_*$ as shown in Fig.~\ref{B8B9s}. From this analysis we can find whether point $B_9$
can be a late time accelerated attractor or a saddle point,
depending on the values of $\xi$, $s_*$ and $dg(s_*)$.
In what follows we will consider again the two examples of scalar field potential analysed in Sec.~\ref{sec:sqare_kc}.
This will allow us to investigate the dynamics of these specific models in more details.

\subsection{Example 1: $V=V_0 \sinh^{-\alpha}(\lambda\phi)$}

For this potential we have again
\begin{equation}\label{eq:002}
g(s)=\frac{1}{\alpha}-\frac{\alpha \lambda^2}{s^2} \,.
\end{equation}
We can now complete the stability analysis for point~$B_4$. The
attractive nature of this point is confirmed numerically by
plotting phase space projections on $x,~y,~s$ separately: for example from
Figs.~\ref{fig4}-\ref{fig6}, we see that trajectories near point
$B_4$ approaches the values $x=-\xi\sqrt{\frac{1}{\xi^2+2}} \simeq 0.58$,
$y=\sqrt{\frac{2}{\xi^2+2}} \simeq 0.82$ and $s=0$ as $N\rightarrow\infty$, implying that point $B_4$ is a late time attractor.
Moreover we have checked that different values of $\xi$ and $\alpha$ do not change this result, unless both $\alpha$ and $\xi$ are positive in which case $B_4$ is a saddle and the late time attractor is $B_5$.
As mentioned above point $B_5$ is independent of the form of the potential for its existence, but it depends on it for its stability.
In this case it is a late time attractor if $3\alpha\xi^2>\alpha^3 \lambda^2$ and a saddle if $3\alpha\xi^2<\alpha^3 \lambda^2$.
Point~$B_6$ is an unstable node if $\alpha\lambda<\sqrt{6}$ and $\alpha<0$, otherwise it is saddle.
Similarly point~$B_7$ is an unstable node for $\alpha\lambda>-\sqrt{6}$ and $\alpha<0$, otherwise it is a saddle point. The regions of existence
and stability in the $(\xi, \lambda)$ parameter space (for $\alpha=4$) of points $B_8$ and $B_9$ are given in Fig.~\ref{B8B9}. Point $B_8$ is stable whenever $\alpha>0$ and saddle if $\alpha<0$; while point $B_9$ is a late time accelerated attractor for some values of the parameters $\xi$ and $\lambda$, as outlined in Fig.~\ref{B8B9}.

The phase space dynamics projected onto the $s=s_*$ plane has been plotted in Fig.~\ref{spB9} for some specific values of the parameters.
Again the dynamics on this plane is similar to the one obtained with an exponential potential (cf.~\cite{Nicola}), though we stress again that now we have to deal with a further dimension.
Fortunately we have numerically checked that the $s=s_*$ plane attracts all other nearby trajectories in the phase space and thus the
late-time evolution of the universe can be almost completely captured by the dynamics on this plane.

\begin{figure}
\centering
\includegraphics[width=8cm,height=6cm]{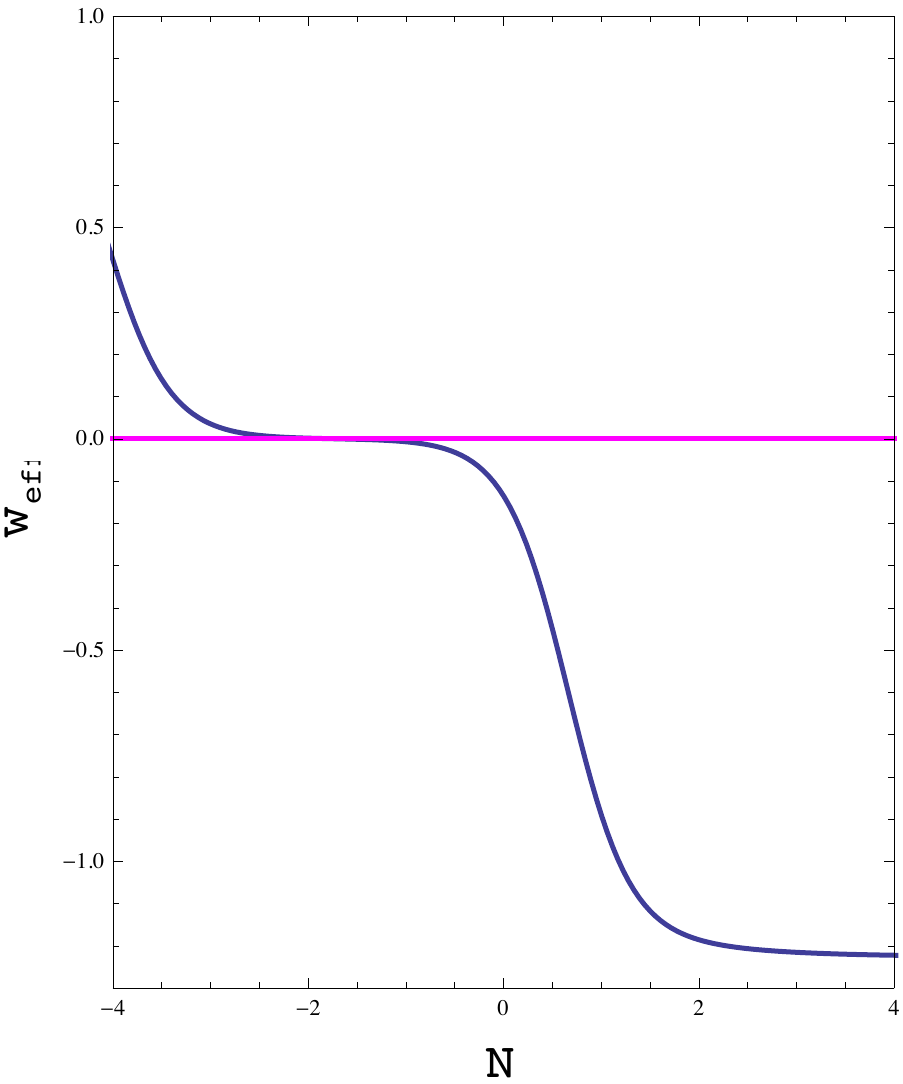}
\caption{Plot of $w_{\rm eff}$ versus $N$ with the potential  $V=V_0 \sinh^{-\alpha}(\lambda\phi)$. Here we have chosen $w=0,~\xi=1,~\alpha=-4,~\lambda=\frac{1}{4}$, showing a transition from matter domination to the phantom regime.}
\label{fig7}
\end{figure}

From Fig.~\ref{spB9} we can realise that within this model one can attain not only general dark matter to acceleration transitions,
 but more specifically dark matter to quintessence-like domination ($B_1$ to $B_9$ in (c)), dark matter to de Sitter acceleration ($B_1$ to $B_4$ or $B_5$,
  respectively in (d) and (e))
, dark matter to phantom domination ($B_1$ to $B_9$ in (a)) and also scaling solutions (point $B_8$ in (b)).
The more interesting scenario, since it cannot be obtained with a canonical scalar field (quintessence), consists in the possibility of dynamically reaching the phantom region.
In fact for specific choices of the parameters, the scalar field characterizes a quintom scenario by crossing the phantom barrier ($w_{\rm eff}=-1$).
This is better exposed in Fig.~\ref{fig7}, which shows also that the universe undergoes a matter to phantom energy transition with an early-time long lasting period of matter domination, as required in order for all cosmic structures to form. 

\subsection{Example 2: $V=\frac{M^{4+n}}{\phi^n}$}

\begin{figure}[t]
\centering
\includegraphics[width=6cm,height=5cm]{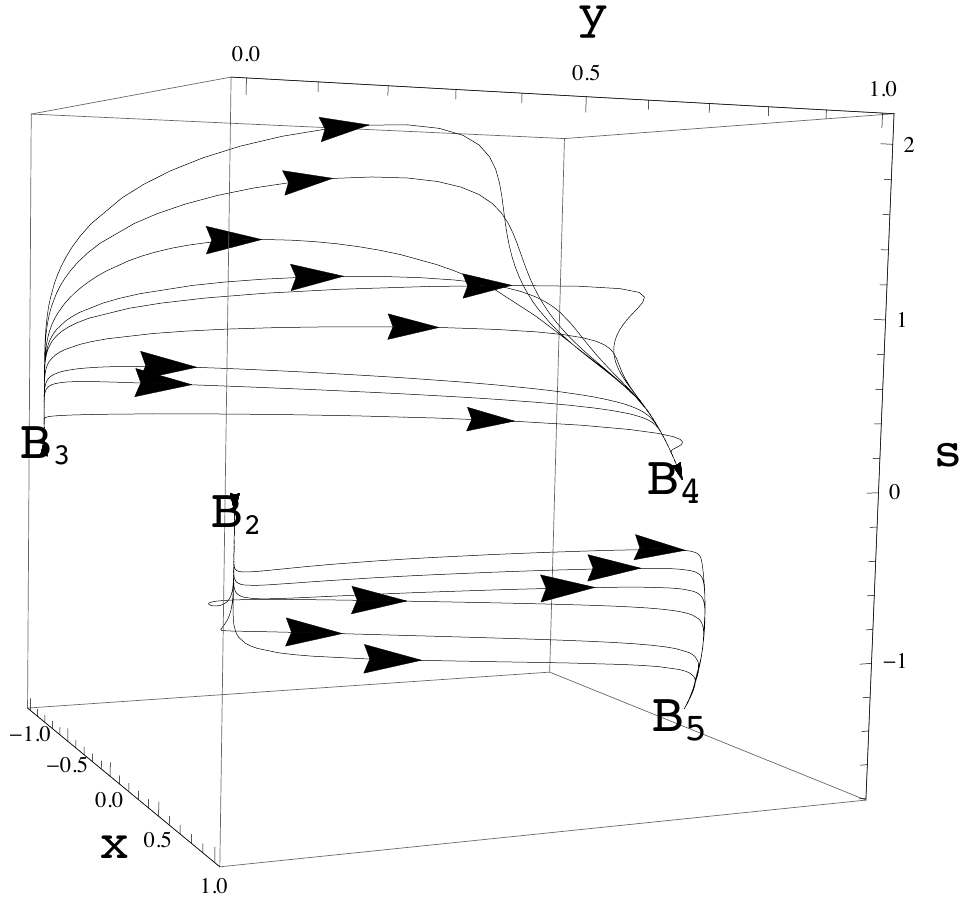}
\caption{3D phase space trajectories of the system (\ref{eq38})-(\ref{eq40}) with $V=\frac{M^{4+n}}{\phi^n}$ and $w=0$.  The parameters chosen here are $n=4$ and $\xi=1$.}\label{contour_B4_B5_3D}
\end{figure}

In this case we have again simply
\begin{equation}\label{eq:003}
g(s)=\frac{1}{n} \,.
\end{equation}
This means that there are no solutions of $g(s) = 0$ and consequently points $B_6-B_9$ do not exist.
So we have only up to five critical points in the phase space: a saddle matter dominated point $B_1$; two stiff matter dominated points $B_2$ and $B_3$,
 which behave as unstable nodes; a scalar field dominated point $B_4$, corresponding to a late time accelerated attractor;
  another scalar field dominated point~$B_5$, corresponding to a late time attractor for $n>0$ and to a saddle if $n<0$.

In this scenario we can have two late time attractors if $n>0$, namely points $B_4$ and $B_5$.
In the $n<0$ case, point $B_5$ is a saddle and the unique late time attractor is point $B_4$.
It is interesting to analyse the dynamics in the $n>0$ situation.
For this purpose we have chosen specific values of the parameters and drawn some phase space trajectories in Fig.~\ref{contour_B4_B5_3D}.
These clearly show that, depending on initial conditions (specifically depending on starting from point $B_2$ or $B_3$), the universe can either finish in point $B_4$ or point $B_5$.
In both cases however a de Sitter solution is attained since $w_{\rm eff}=-1$, and thus the late time accelerated expansion of the universe can be described consistently.
Moreover some trajectories, though not among the ones shown in Fig.~\ref{contour_B4_B5_3D}, can actually describe a matter to dark energy transition since point $B_1$ represents a saddle matter dominated critical point.
Requiring an suitable level of fine tuning of initial conditions, the observed evolution history of the universe can thus be reproduced also in this model.

\section{Conclusions} 
\label{sec:conclusions}

This work has been devoted to study the late-time cosmology arising from two scalar field models of dark energy, where square and square root kinetic corrections to the canonical Lagrangian appear.
The scope was to extend to arbitrary potentials the dynamical systems analysis performed in \cite{Nicola} uniquely for the exponential scalar field potential.

From our results we have found that interesting cosmological implications can arise from these scalar field models.
Several physically useful solutions can in fact be found: de Sitter solutions, quintessence-like solutions, matter dominated solutions, scaling solutions and phantom dominated solutions.
A matter to dark energy transition, effectively describing the cosmic evolution that one reconstructs from observations at the background level, can be achieved in all the models analysed, with a possible relaxing on the fine tuning of initial conditions in at least one of them (see Sec.~\ref{subsec:square_2}).
It would be interesting to investigate the cosmological perturbations of these extended scalar field models, in order to find possible observational signatures that can be compared with astronomical data.
However such analysis lies outside the scopes of the present work and will thus be left for future studies.

Moreover within these models one can achieve not only late-time accelerated attractors, but also a possible dynamical crossing of the phantom barrier, which is slightly favoured by present observations \cite{Xia:2013dea,Novosyadlyj:2013nya} and it cannot be obtained with a canonical scalar field (quintessence).
Unfortunately such cosmic evolutions are generically hunted by instabilities at the perturbation level when driven by scalar fields \cite{Vikman:2004dc,Zhao:2005vj,Caldwell:2005ai}, and the models investigated in this work make no exceptions.
Nevertheless there might be some other descriptions of nature that lead to the same cosmological dynamics at the background level, making the analysis of this paper physically relevant.
This is exactly what happens to the models considered in this work when an exponential potential is selected \cite{Nicola}.
In fact, as shown in \cite{Boehmer:2015sha}, an identical cosmological background dynamics arises if the scalar field, instead of being corrected by kinetic terms, is assumed to interact with matter, particularly in the case of the so-called Scalar-Fluid theories \cite{Boehmer:2015kta,Boehmer:2015sha,Koivisto:2015qua,Boehmer:2015ina,Brax:2015fcf}.

In conclusion the analysis presented in this work shows that the results obtained in \cite{Nicola} with an exponential potential, can nicely be re-obtained and extended with other scalar field potentials, in a similar fashion to what happens with quintessence (see e.g.~\cite{TamaniniPhDthesis}).
The generalisation to other scalar field potentials is important not only from a mathematical point of view, but also to better connect these phenomenological models with the low-energy limit of more fundamental high energy theories.
Furthermore if future cosmological probes will confirm the phantom nature of dark energy, then it will be no more possible for us to rely upon neither the cosmological constant nor the quintessence field.
In such a scenario any alternative model capable of matching the observational data in a consistent way will become useful, irrespectively of its phenomenological nature.


\acknowledgments

J.D.~is thankful to IUCAA for warm hospitality and
facility of doing research works.
N.T.~acknowledge support from the Labex P2IO and the Enhanced Eurotalents Programme.

\appendix

\section*{Appendix} 
\label{sec:app}

In this appendix we provide the eigenvalues of $A_4$ and its corresponding $w_{\rm eff}$ assuming $w=0$: \\ $ $ \\
$E_1=\frac{1}{12\,\Xi ^{2/3} \left( 4\,\xi+
1 \right)}\left[-648\,6^{1/3}s_*^{4}\xi-108\,6^{1/6}
{\Xi^{1/3}}s_*^{3}\xi\right.$\\$\left.+6^{1/3}\,30\,s_*^{4}+7776\,
s_*^{2}\,6^{1/3}\,{\xi}^{2}+21\,6^{1/6}\,{\Xi^{1/3}}s_*^{3}+108\,\Delta\,6^{1/3}\,s_*\xi+6^{1/3}\,1296\,{
s_*}^{2}\xi+12\,{\Xi}^{2/3}s_*^{2}-432\,{\Xi^{1/3}}6^{1/6}\,s_*\,\xi\right.$\\$\left.+27\,\Delta\,6^{1/3}\,s_*-6^{1/3}\,162\,{s_*}^{2}-144\,{\Xi}^{2/3}\xi-108\,{\Xi^{1/3}}\,6^{1/6}\,
s_*-36\,{\Xi}^{2/3}+ \left(-1296\, \sqrt{6}{\Xi^{1/3}}\Delta\,{
s_*}^{4}\xi-93312\, \sqrt{6}{\Xi^{1/3}}\Delta\,{s_*}^{2}{
\xi}^{2}\right.\right.$\\$\left.\left.-46656\, \sqrt{6}{\Xi^{1/3}}\Delta\,{s_*}^{2}\xi+6^{1/3}\,288\,
 {\Xi}^{2/3}\Delta\,{s_*}^{3}\xi+{6}^{2/3}\,417\,{s_*}^{8}+
{6}^{2/3}\,46656\,{s_*}^{2}-{6}^{2/3}\,5832\,{s_*}^{4}-96\,{\Xi}^{
4/3}{s_*}^{4}+576\,{\Xi}^{4/3}{s_*}^{2}\right.\right.$\\$\left.\left.+1296\,{\Xi}^{2/3}
6^{1/3}\,{s_*}^{6}{\xi}^{2}-2232\,{\Xi}^{2/3}6^{1/3}{
s_*}^{6}\xi-25920\,{6}^{2/3}\Delta\,{s_*}^{5}{\xi}^{2}+31104\,
{\Xi}^{2/3}\,6^{1/3}\,{s_*}^{4}{\xi}^{2}-3744\,{6}^{2/3}\Delta\,{
s_*}^{5}\xi\right.\right.$\\$\left.\left.+186624\,{6}^{2/3}\Delta\,{s_*}^{3}{\xi}^{3}+4032\,
{\Xi}^{2/3}\,6^{1/3}\,{s_*}^{4}\xi+36288\,{6}^{2/3}\Delta\,{
s_*}^{3}{\xi}^{2}+20736\,{\Xi}^{2/3}\,6^{1/3}\,{s_*}^{2}{\xi}
^{2}-16848\,{6}^{2/3}\Delta\,{s_*}^{3}\xi\right.\right.$\\$\left.\left.+72\,{\Xi}^{2/3}\,6^{1/3}\,\Delta\,{s_*}^{3}+10368\,{\Xi}^{2/3}\,6^{1/3}\,{s_*}^{2}
\xi-23328\, \sqrt{6}\Delta\,{s_*}^{4}{\xi}^{2}+1134\, 6^{1/3}\,{\Xi}\Delta\,{s_*}^{4}-5832\, \sqrt{6}{\Xi^{1/3}}
\Delta\,{s_*}^{2}\right.\right.$\\$\left.\left.+139968\, \sqrt{6}{\Xi^{1/3}}{s_*}^{7}{\xi
}^{2}-33696\, \sqrt{6}{\Xi^{1/3}}{s_*}^{7}\xi-1679616\, \sqrt{6
}{\Xi^{1/3}}{s_*}^{5}{\xi}^{3}+606528\, \sqrt{6}{\Xi^{1/3}}{
s_*}^{5}{\xi}^{2}+203472\, \sqrt{6}{\Xi^{1/3}}{s_*}^{5}\xi\right.\right.$\\$\left.\left.-
6718464\, \sqrt{6}{\Xi^{1/3}}{s_*}^{3}{\xi}^{3}-2799360\,
 \sqrt{6}{\Xi^{1/3}}{s_*}^{3}{\xi}^{2}-139968\, \sqrt{6}{\Xi^{1/3}}{s_*}^{3}\xi+34992\, \sqrt{6}{\Xi^{1/3}}{s_*}^{3}-
2772\,{6}^{2/3}{s_*}^{6}{\Xi^{1/3}}\right.\right.$\\$\left.\left.-18324\,{6}^{2/3}{s_*}^{
8}\xi-936\,{\Xi}^{2/3}6^{1/3}{s_*}^{4}+2304\,{\Xi}^{4/3}{
s_*}^{2}\xi-13284\, \sqrt{6}{\Xi^{1/3}}{s_*}^{5}-2426112\,{
6}^{2/3}{s_*}^{6}{\xi}^{3}+684\,{6}^{2/3}\Delta\,{s_*}^{5}\right.\right.$\\$\left.\left.+
13436928\,{6}^{2/3}{s_*}^{4}{\xi}^{4}+129\,{\Xi}^{4/3}{s_*}^{4
}\xi-746496\,{6}^{2/3}{s_*}^{4}{\xi}^{2}+559872\,{6}^{2/3}{s_*
}^{2}\xi+1296\,{\Xi}^{2/3}6^{1/3}{s_*}^{2}+2239488\,{6}^{2/3}
{s_*}^{2}{\xi}^{2}\right.\right.$\\$\left.\left.-3564\,{6}^{2/3}\Delta\,{s_*}^{3}+1260\,
 \sqrt{6}{\Xi^{1/3}}{s_*}^{7}+2985984\,{6}^{2/3}{s_*}^{2}{
\xi}^{3}-233280\,{6}^{2/3}{s_*}^{4}\xi+46656\,{6}^{2/3}{s_*}^{
8}{\xi}^{3}+128304\,{6}^{2/3}{s_*}^{8}{\xi}^{2}\right)^{1/2}\right]$
\\[1.5ex]
$E_2=\frac{1}{12\,\Xi ^{2/3} \left( 4\,\xi+
1 \right)}\left[-648\,6^{1/3}s_*^{4}\xi-108\,6^{1/6}
\sqrt [3]{\Xi}s_*^{3}\xi\right.$\\$\left.+6^{1/3}\,30\,s_*^{4}+7776\,
s_*^{2}\,6^{1/3}\,{\xi}^{2}+21\,\sqrt [3]\,6^{1/6}\,{\Xi}s_*^{3}+108\,\Delta\,6^{1/3}\,s_*\xi+6^{1/3}\,1296\,{
s_*}^{2}\xi+12\,{\Xi}^{2/3}s_*^{2}-432\,{\Xi^{1/3}}6^{1/6}\,s_*\,\xi\right.$\\$\left.+27\,\Delta\,6^{1/3}\,s_*-6^{1/3}\,162\,{s_*}^{2}-144\,{\Xi}^{2/3}\xi-108\,{\Xi^{1/3}}\,6^{1/6}\,
s_*-36\,{\Xi}^{2/3}- \left(-1296\, \sqrt{6}{\Xi^{1/3}}\Delta\,{
s_*}^{4}\xi-93312\, \sqrt{6}{\Xi^{1/3}}\Delta\,{s_*}^{2}{
\xi}^{2}\right.\right.$\\$\left.\left.-46656\, \sqrt{6}{\Xi^{1/3}}\Delta\,{s_*}^{2}\xi+6^{1/3}\,288\,
 {\Xi}^{2/3}\Delta\,{s_*}^{3}\xi+{6}^{2/3}\,417\,{s_*}^{8}+
{6}^{2/3}\,46656\,{s_*}^{2}-{6}^{2/3}\,5832\,{s_*}^{4}-96\,{\Xi}^{
4/3}{s_*}^{4}+576\,{\Xi}^{4/3}{s_*}^{2}\right.\right.$\\$\left.\left.+1296\,{\Xi}^{2/3}
6^{1/3}\,{s_*}^{6}{\xi}^{2}-2232\,{\Xi}^{2/3}6^{1/3}{
s_*}^{6}\xi-25920\,{6}^{2/3}\Delta\,{s_*}^{5}{\xi}^{2}+31104\,
{\Xi}^{2/3}\,6^{1/3}\,{s_*}^{4}{\xi}^{2}-3744\,{6}^{2/3}\Delta\,{
s_*}^{5}\xi\right.\right.$\\$\left.\left.+186624\,{6}^{2/3}\Delta\,{s_*}^{3}{\xi}^{3}+4032\,
{\Xi}^{2/3}\,6^{1/3}\,{s_*}^{4}\xi+36288\,{6}^{2/3}\Delta\,{
s_*}^{3}{\xi}^{2}+20736\,{\Xi}^{2/3}\,6^{1/3}\,{s_*}^{2}{\xi}
^{2}-16848\,{6}^{2/3}\Delta\,{s_*}^{3}\xi\right.\right.$\\$\left.\left.+72\,{\Xi}^{2/3}\,6^{1/3}\,\Delta\,{s_*}^{3}+10368\,{\Xi}^{2/3}\,6^{1/3}\,{s_*}^{2}
\xi-23328\, \sqrt{6}\Delta\,{s_*}^{4}{\xi}^{2}+1134\, 6^{1/3}\,{\Xi}\Delta\,{s_*}^{4}-5832\, \sqrt{6}{\Xi^{1/3}}
\Delta\,{s_*}^{2}\right.\right.$\\$\left.\left.+139968\, \sqrt{6}{\Xi^{1/3}}{s_*}^{7}{\xi
}^{2}-33696\, \sqrt{6}{\Xi^{1/3}}{s_*}^{7}\xi-1679616\, \sqrt{6
}{\Xi^{1/3}}{s_*}^{5}{\xi}^{3}+606528\, \sqrt{6}{\Xi^{1/3}}{
s_*}^{5}{\xi}^{2}+203472\, \sqrt{6}{\Xi^{1/3}}{s_*}^{5}\xi\right.\right.$\\$\left.\left.-
6718464\, \sqrt{6}{\Xi^{1/3}}{s_*}^{3}{\xi}^{3}-2799360\,
 \sqrt{6}{\Xi^{1/3}}{s_*}^{3}{\xi}^{2}-139968\, \sqrt{6}{\Xi^{1/3}}{s_*}^{3}\xi+34992\, \sqrt{6}{\Xi^{1/3}}{s_*}^{3}-
2772\,{6}^{2/3}{s_*}^{6}{\Xi^{1/3}}\right.\right.$\\$\left.\left.-18324\,{6}^{2/3}{s_*}^{
8}\xi-936\,{\Xi}^{2/3}6^{1/3}{s_*}^{4}+2304\,{\Xi}^{4/3}{
s_*}^{2}\xi-13284\, \sqrt{6}{\Xi^{1/3}}{s_*}^{5}-2426112\,{
6}^{2/3}{s_*}^{6}{\xi}^{3}+684\,{6}^{2/3}\Delta\,{s_*}^{5}\right.\right.$\\$\left.\left.+
13436928\,{6}^{2/3}{s_*}^{4}{\xi}^{4}+129\,{\Xi}^{4/3}{s_*}^{4
}\xi-746496\,{6}^{2/3}{s_*}^{4}{\xi}^{2}+559872\,{6}^{2/3}{s_*
}^{2}\xi+1296\,{\Xi}^{2/3}6^{1/3}{s_*}^{2}+2239488\,{6}^{2/3}
{s_*}^{2}{\xi}^{2}\right.\right.$\\$\left.\left.-3564\,{6}^{2/3}\Delta\,{s_*}^{3}+1260\,
 \sqrt{6}{\Xi^{1/3}}{s_*}^{7}+2985984\,{6}^{2/3}{s_*}^{2}{
\xi}^{3}-233280\,{6}^{2/3}{s_*}^{4}\xi+46656\,{6}^{2/3}{s_*}^{
8}{\xi}^{3}+128304\,{6}^{2/3}{s_*}^{8}{\xi}^{2}\right)^{1/2}\right]$\\[1.5ex]

$E_3=-\sqrt{6} x_4 s_*^2 dg(s_*)$\\[1.5ex]
$w_{\rm eff}=\frac {\sqrt {6}{s_*}^{3}\mu+24\,{s_*}^{2}\xi\,{\mu}^
{2}-28\,\sqrt {6}s_*\,\xi\,\mu-2\,{s_*}^{2}{\eta}^{2}-3\,
\sqrt {6}s_*\,\mu+24\,\xi\,{\mu}^{2}-4\,{s_*}^{2}+6\,{\mu}^
{2}+48\,\xi+12}{3 \left( 4\,\xi+1 \right)  \left( -\sqrt {6}s_*\,
\mu+2\,{\mu}^{2}+4 \right)}$ \\ $ $ \\

Here we have used:
\begin{eqnarray}
\mu&=&-{\frac {36\,{6}^{2/3}{s_*}^{2}\xi-7\,{6}^{2/3}{s_*}^{2}
+144\,{6}^{2/3}\xi-{6^{1/3}}{\Xi}^{2/3}-4\,\sqrt {6}s_*\,{\Xi^{1/3}}+36\,{6}^{2/3}}{18 \left( 4\,\xi+1 \right) {\Xi^{1/3}}}}\nonumber\\
\Delta&=&\sqrt {36\,{s_*}^{6}\xi-3\,{s_*}^{6}-432\,{s_*}^{4}\xi+52
\,{s_*}^{4}+5184\,{s_*}^{2}{\xi}^{2}+864\,{s_*}^{2}\xi-300
\,{s_*}^{2}+2304\,\xi+576}\nonumber\\
\Xi&=&-216\,\sqrt {6}{s_*}^{3}\xi+10\,\sqrt {6}{s_*}^{3}+2592\,
\sqrt {6}s_*\,{\xi}^{2}+36\,\sqrt {6}\Delta\,\xi+432\,\sqrt {6}
s_*\,\xi+9\,\sqrt {6}\Delta-54\,\sqrt {6}s_*\nonumber
\end{eqnarray}

\end{document}